\documentstyle[epsfig,a4,12pt]{article}
\begin{document}

\newcommand{\x}{\cdot}
\newcommand{\ra}{\rightarrow}
\newcommand{\pom}{\mbox{${\rm \cal P}$omeron}}
\newcommand{\flux}{\mbox{$F_{{\cal P}/p}(t, \xi)$}}
\newcommand{\ap}{\mbox{$\bar{p}$}}
\newcommand{\pap}{\mbox{$p \bar{p}$}}
\newcommand{\SPS}{\mbox{S\pap S}}
\newcommand{\xp}{\mbox{$x_{p}$}}
\newcommand{\sumet}{\mbox{$\Sigma E_t$}}
\newcommand{\mpi}{\mbox{${m_\pi}$}}
\newcommand{\rs}{\mbox{$\sqrt{s}$}}
\newcommand{\rsp}{\mbox{$\sqrt{s'}$}}
\newcommand{\rsps}{\mbox{$\sqrt{s} = 630 $ GeV}}
\newcommand{\lum}{\mbox{$\int {\cal L} {dt}$}}
\newcommand{\T}{\mbox{$t$}}
\newcommand{\abt}{\mbox{${|t|}$}}
\newcommand{\di}{\mbox{d}}
\newcommand{\HS}{\mbox{$xG(x)=6x(1-x)^1$}}
\newcommand{\sigdifjets}{\mbox{$\sigma_{sd}^{jets}$}}
\newcommand{\sigpomjets}{\mbox{$\sigma_{{\cal P}p}^{jets}$}}
\newcommand{\sigdiftot}{\mbox{$\sigma_{sd}^{total}$}}
\newcommand{\sigpomtot}{\mbox{$\sigma_{{\cal P}p}^{total}$}}
\newcommand{\sigpomzero}{\mbox{$\sigma_{{\cal P}p}^o$}}
\newcommand{\dsig}{\mbox{${d^2 \sigma }\over{d \xi dt}$}}
\newcommand{\alamb}{\mbox{$\overline{\Lambda}^{\circ}$}}
\newcommand{\lamb}{\mbox{$\Lambda^{\circ}$}} 
\newcommand{\peetee}{\mbox{${p_t}$}}
\newcommand{\PRET}{\mbox{\Proton-\sumet}}
%
\begin{titlepage}
\begin{center}
{\large   }
{\large EUROPEAN ORGANIZATION FOR NUCLEAR RESEARCH}
\end{center}
\begin{flushright}  {
9 December, 1997
}
\end{flushright}
 
{
\Large
\centerline{\bf Measurements of Single Diffraction at {\boldmath \rsps };}
\centerline{\bf Evidence for a Non-Linear {\boldmath $\alpha(t)$} of the
{\boldmath \pom}}
\normalsize
}
\vspace{4 ex}
\begin{center}
A.~Brandt$^{1}$, 
S.~Erhan$^{a}$,
A.~Kuzucu$^{2}$, 
D.~Lynn$^{3}$, 
M.~Medinnis$^{4}$,\\
N.~Ozdes$^{2}$, 
P.E.~Schlein$^{b}$                       
M.T.~Zeyrek$^{5}$, 
J.G.~Zweizig$^{6}$\\
University of California$^{*}$, Los Angeles, California 90095, USA. \\
\vspace{3 ex}
J.B. Cheze, J. Zsembery \\
Centre d'Etudes Nucleaires-Saclay, 91191 Gif-sur-Yvette, France.
\end{center}
\vspace{2 ex}

\centerline{(UA8 Collaboration)}

\begin{abstract}
We report measurements of the inclusive differential cross section
for the single-diffractive reactions:
$p_i + \ap \, \ra \, p_f + X$ \, and \, 
$p + \ap _i  \, \ra \, X + \ap_f$
at \rsps , in the momentum transfer range, $0.8 < -t < 2.0$~GeV$^2$
and final state Feynman-$\xp > 0.90$. 
Based on the assumption of factorization,
several new features of the \pom\ Flux Factor are determined from
simultaneous fits to our UA8 data and lower energy data from the CHLM 
collaboration at the CERN-Intersecting Storage Rings.
 
Prominent among these is that the effective
\pom\ Regge trajectory requires a term 
quadratic in \T , with coefficient, $\alpha '' = 0.079 \pm 0.012$~GeV$^{-4}$.
We also show that
the data require a \pom -proton cross section that first decreases 
with increasing diffractive mass (corresponding to the 
${\cal P}{\cal P }{\cal R }$ term in the triple-Regge expansion) 
and then increases at larger mass (the ${\cal P }{\cal P}{\cal P}$ term), 
similar to real particle total cross sections. 
We measure the product, 
$K \sigma _0 = 0.72 \pm 0.10$~mb~GeV$^{-2}$, where $K$ is the 
normalization constant of the \pom\ Flux Factor in the proton and $\sigma _0$
is the scale constant in the \pom -proton total cross section.
Finally, we report the occurence of ``beam jets'' in the \pom\ direction in
the rest frame of the diffractive system.
\end{abstract}
\begin{center}
Nuclear Physics B (in press - 1997)\\
\end{center}
\vspace{1 ex}
\rule[.5ex]{16cm}{.02cm}
$^{*}$ Supported by U.S. National Science Foundation
Grant PHY94-23142 \\
$^{a}$ Email: samim.erhan@cern.ch \\
$^{b}$ Email: peter.schlein@cern.ch \\
$^{1}$ Now at Fermi National Accelerator Laboratory, Batavia, Illinois, 
U.S.A. \\ 
$^{2}$ Visitor from Cukurova University, Adana, Turkey; also supported by 
ICSC - World Lab.\\
$^{3}$ Now at Brookhaven National Laboratory, Upton, Long Island, NY \\
$^{4}$ Present address: DESY, Zeuthen, Germany \\
$^{5}$ Visitor from Middle East Tech. Univ., Ankara, Turkey; supported by Tubitak. \\
$^{6}$ Present address: DESY, Hamburg, Germany  \\

\end{titlepage}      

\pagebreak
\setlength{\oddsidemargin}{0 cm}
\setlength{\evensidemargin}{0 cm}
\setlength{\topmargin}{0.5 cm}
\setlength{\textheight}{22 cm}
\setlength{\textwidth}{16 cm}
\setcounter{totalnumber}{20}
\clearpage\mbox{}\clearpage
\pagestyle{plain}
\setcounter{page}{1}

\tableofcontents

\pagebreak
 
\section{Introduction}
\label{intro}
\indent 

We report a study of the single-diffractive reaction:
\begin{equation}
p_i \, + \, \ap\ \,  \ra \, p_f \,+\, X 
\label{eq:dif}
\end{equation}
and its charge conjugate
at the CERN  S\pap S-Collider with \rsps , where 
$p_i$ and $p_f$ are, respectively, the inital and final state proton momenta.

\begin{figure}
\begin{center}
\mbox{\epsfig{file=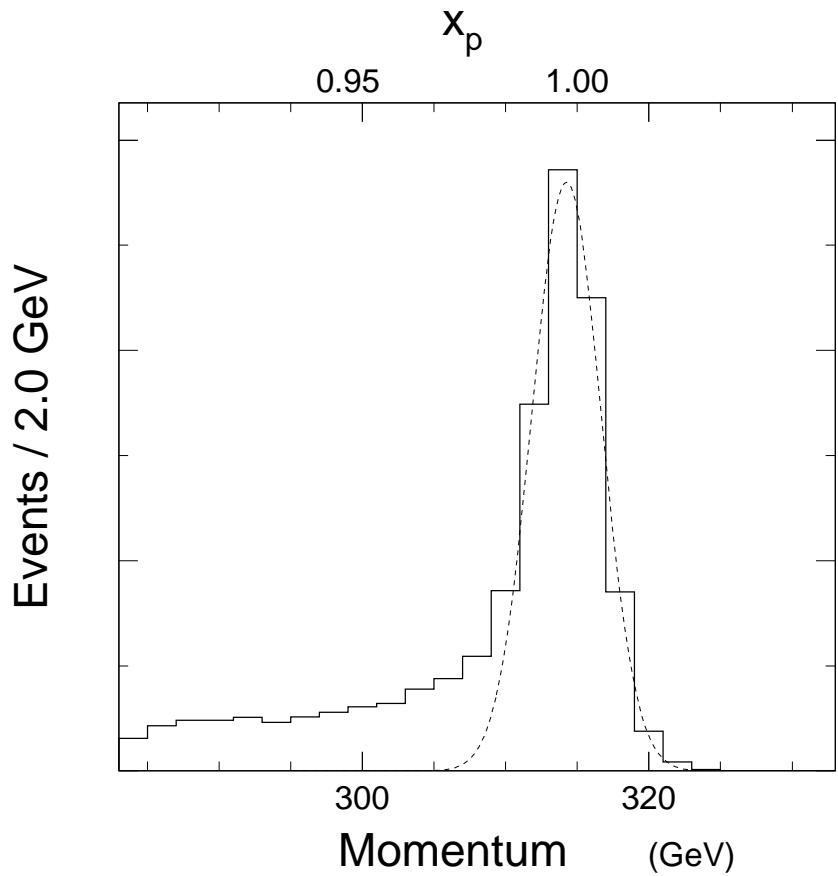,width=12.7cm}}
\end{center}
\caption[]{
A raw \ap\ momentum distribution measured in this experiment, 
summed over all t.
The dashed curve is a Gaussian resolution  which mirrors
the high side of the distribution, and shows that the 
resolution smearing of the peak is mainly confined to
\xp\ values larger than about 0.97.
}
\label{fig:mumom}
\end{figure}

\begin{figure}
\begin{center}
\mbox{\epsfig{file=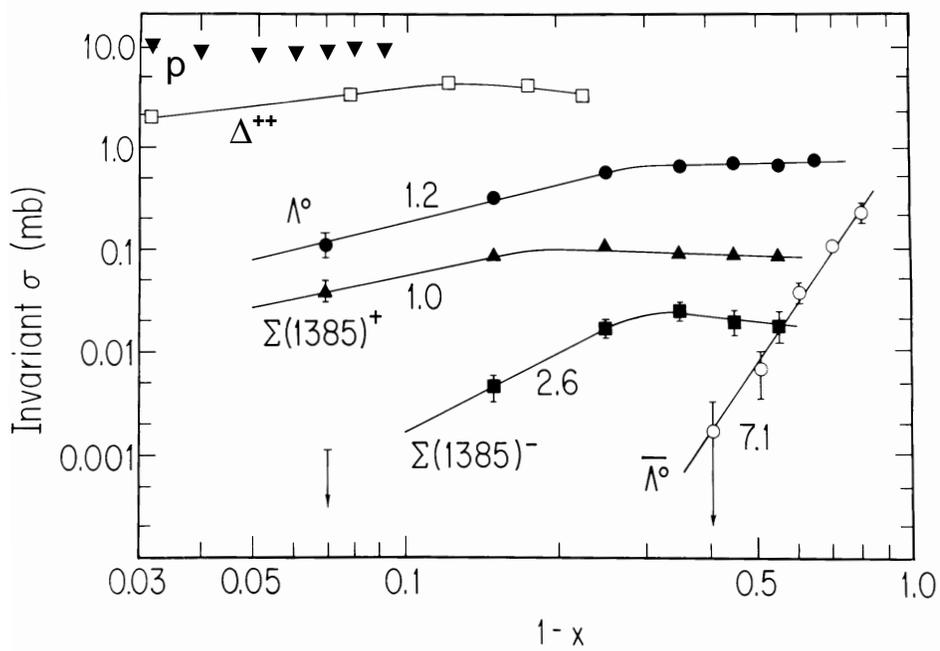,width=12.7cm}}
\end{center}
\caption[]{
Invariant cross sections for (inclusive) forward 
baryons\cite{erhan} in proton-proton interactions
at the CERN Intersecting Storage Rings, integrated over
transverse momentum, ($x$/$\pi$)$d \sigma$/$dx$ vs. $\xi = 1 - x$. 
$\sigma$ includes
a factor of 2, to account for production in both hemispheres. 
$\Delta ^{++}$ points are from Ref.~\protect\cite{lockman}.
The proton points are estimates from UA8.
The numbers labeling each set of data points is the exponent of the 
straight line $(1 - x)^n$
}
\label{fig:baryons}
\end{figure}

One of the most remarkable phenomena in strong interaction physics occurs
in reactions of this type, which have beam-like particles in the final state.
It has been known for many years\cite{difreview} that the inclusive 
spectra of final-state particles which have the identity
of the beam particle increase rapidly as Feynman-$x$ of the particle 
approachs unity,
in striking contrast with the (1-$x$)$^n$ (with $n>0$) type of 
falling spectra observed for all other particle types.
Figure~\ref{fig:mumom} shows such a spectrum for final state 
antiprotons\footnote{Since we henceforth discuss protons (antiprotons),
we use the notation, \xp .}
in the charge conjugate of React.~\ref{eq:dif}. 
For $\xp < 0.97$, the distribution is rather flat
but with a much larger cross section than for other baryons. 
For example, this is shown in Fig.~\ref{fig:baryons} 
for many types of final-state baryons\cite{erhan,lockman}.
The exponent which labels each data set in the figure depends inversely 
on the number of beam valence quarks in the final-state baryon; 
thus, the relatively flat proton distribution is consistent with
the expectation that the final-state proton contains all the
valence quarks of the beam proton. 

\begin{figure}
\begin{center}
\mbox{\epsfig{file=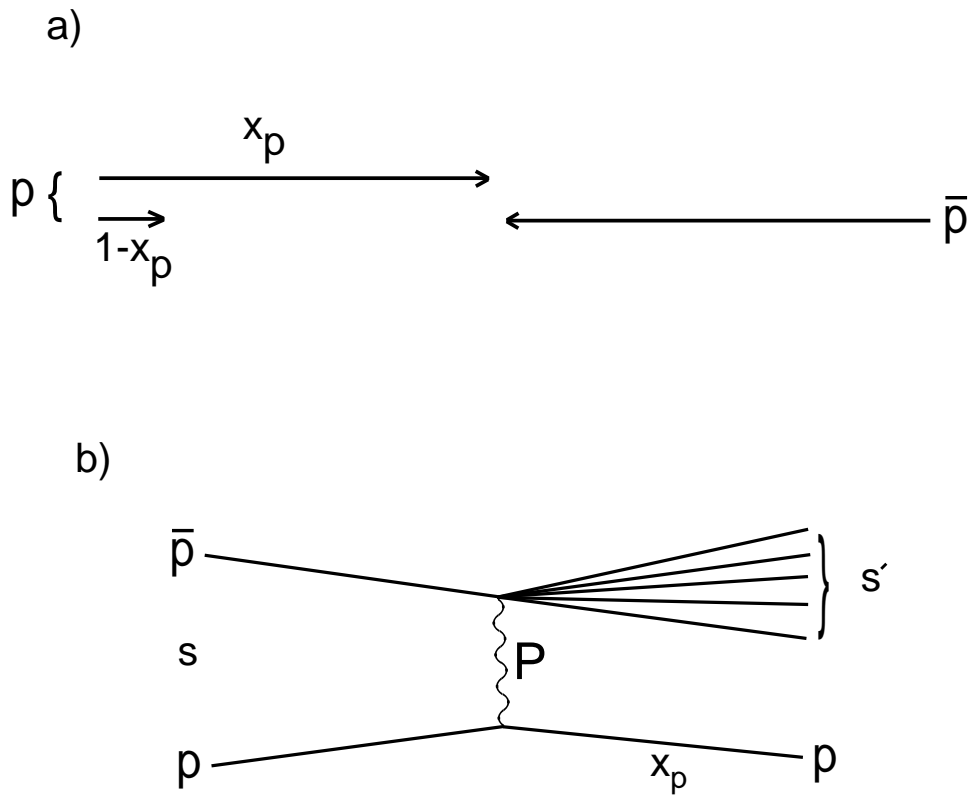,width=12.7cm}}
\end{center}
\caption[]{
(a) Incident \pom\ with beam momentum fraction $\xi = 1 - \xp$ interacts 
with the incident \ap .
(b) Characteristic t-channel 
diagram for diffractive processes in which a \pom\ is exchanged. $s$ and $s'$ 
are kinematically related by: $s' = (1-\xp)s = \xi s$.
The observed spectator proton has Feynman-\xp .}
\label{fig:difdiag}
\end{figure}

In React.~\ref{eq:dif}, an observed rapidity gap between $p_f$ and $X$ in
the final state signifies that the entire residual momentum of the proton 
(with beam fraction, $\xi = 1-\xp$) participates in the interaction 
between it and the second beam particle.
This effect has been described\cite{difreview} in terms of an exchange of 
the \pom\ Regge trajectory, which embodies the idea of ``factorization''.
The interacting vectors of the \pap\ system shown in Fig.~\ref{fig:difdiag}(a)
can be rearranged in the so-called ``\T -channel exchange diagram'' shown in 
part (b) of the figure. The exchanged entity, with 
beam momentum fraction, $\xi = 1 - \xp$, and 
squared-momentum-transfer\footnote{We use the metric where $t$ is negative in
the physical region}, $t = (p_i - p_f)^2$, is the \pom .
The upper vertex is a collision with center-of-mass energy \rsp , 
between the \ap\ and the soft \pom -dominated component of $p_i$.

Since $\xp \sim 1$ is the most likely beam momentum fraction of the final state 
$p_f$ or $\ap _f$, correspondingly the most likely value of the 
\pom 's momentum fraction, $\xi$, is near 
zero\footnote{Because $\xp + \xi = 1$, these are equivalent variables,
and we use them interchangably in this paper.}. 
Thus, a proton beam is essentially a beam
of low momentum \pom s, as depicted in Fig.~\ref{fig:difdiag}(a).
To good approximation, the invariant squared-mass of the $X$-system, $s'$, is 
kinematically related to the total squared-energy in the initial state, $s$,
by the relation: $s' = \xi s$. 
Thus, a measurement of \xp\ tags diffractive events with diffractive mass, 
\rsp .
In high energy collisions, $s'$ can be quite large.
For example, in the experiment reported here, when \xp\ =
0.95 (0.90), we have \rsp\ = 140 (200) GeV.

The concept of factorization and the possibility to transfer 
large amounts of energy-momentum
at high energy led Ingelman and Schlein\cite{is} 
to propose that, in React.~\ref{eq:dif}  
(and, analogously, also in (virtual) $\gamma p$ interactions in $ep$ 
scattering), 
it should be possible to observe hard scattering in the \pom -proton
interaction and to obtain information about the partonic structure of
the \pom . 

Finding and studying jet production in 
React.~\ref{eq:dif} was the original purpose of this experiment and
positive results were reported in three previous UA8 
papers\cite{bonino,brandt,jetsig}.
Jets were observed\cite{bonino} which possessed typical QCD properties,
thus establishing that the \pom\ has a partonic structure.
Subsequent UA8 jet results\cite{brandt} on
React.~\ref{eq:dif} from the 1988--1989 \SPS -Collider runs
showed that the \pom 's structure was hard, and that there appeared to be
a $\delta$-function-like component (in about 30\% of the observed events),
in which the entire momentum of the \pom\ enters into the hard scattering.
The ZEUS\cite{zeus} and H1\cite{h1} $e^+ p$ experiments at HERA 
and the CDF\cite{cdfdijet} and D\O\ \cite{andrewd0} 
\pap\ experiments at Fermilab have
since reported other hard scattering results in diffraction, and 
``hard diffractive scattering" is now 
a well-established component of most high energy experiments.

These hard scattering results, together with the success of
Donnachie and Landshoff\cite{dl_dif} in using factorization and a \pom -photon
analogy to predict React.~\ref{eq:dif} from elastic scattering measurements,
give impetus to the idea that the \pom\ behaves like 
a {\it quasi-real} object inside the proton 
with an effective \pom\ flux factor.

Assuming the validity of factorization, the differential cross section for
React.~\ref{eq:dif} can be expressed as a product of the flux factor
and a \pom -proton scattering cross section, \sigpomtot :
\begin{equation}
{{d^2 \sigma}\over{d \xi dt}} \, \, 
= \, \,  \flux \, \, \x \, \, \sigpomtot(s').
\label{eq:factorization}
\end{equation}
In React.~\ref{eq:dif}, where \sigpomtot\ depends 
relatively weakly on $s'$, the 
shape of the observed $\xp = 1 - \xi$ distribution, before resolution smearing, 
is dominated (at low $|t|$)
by a  $\sim \xi ^{-1}$ factor in \flux , as discussed in Sect.~\ref{combined}.
Contrast this with jet production in React.~\ref{eq:dif}, where 
\sigpomjets\ is zero at low $s'$ and rises 
dramatically with increasing $s'$, such that the $\xp \sim 1$ peak is no
longer visible\cite{jetsig,is}.

In the present paper, we present a detailed study of
inclusive diffraction in React.~\ref{eq:dif}.
The momentum transfer is 
in the range $0.8 \, < \, -t \, < \, 2.0 \: $ GeV$^2$ 
for protons and antiprotons with $\xp > 0.9$.
As discussed in the following section,
the final state $p_f$ or $\ap _f$ is detected in one of four Roman-pot 
spectrometers\cite{ua8hard}, 
while much of the diffractive system, $X$, is detected in the calorimeters  
of the UA2 experiment\cite{ua2hard}, installed in the same interaction region.

In Chapt.~\ref{appar},
the UA8 apparatus and triggers\cite{ua8hard} are described.
We describe the event selection and various corrections applied to the
raw data in Chapt.~\ref{select}. The resulting absolute cross sections
for React.~\ref{eq:dif} are given in Chapt.~\ref{sigmas}.

A combined analysis of the UA8 data and the extensive data on 
React.~\ref{eq:dif} by the CHLM collaboration at the 
CERN Intersecting Storage Rings\cite{albrow} is given 
in Chapt.~\ref{chaptanal}. 
Prominent among these results is that the effective
\pom\ Regge trajectory requires a term 
quadratic in \T , with coefficient, $\alpha '' = 0.079 \pm 0.012$~GeV$^{-4}$.
We also show that
the data require a \pom -proton cross section that first decreases 
with increasing diffractive mass (corresponding to the 
${\cal P}{\cal P }{\cal R }$ term in the triple-Regge expansion) 
and then increases at larger mass (the ${\cal P }{\cal P}{\cal P}$ term), 
similar to real particle total cross sections. 
We measure the product, 
$K \sigma _0 = 0.72 \pm 0.10$~mb/GeV$^2$, where $K$ is the 
normalization constant of the \pom\ flux factor, \flux , 
in the proton and $\sigma _0$
is the scale constant in the \pom -proton total cross section, \sigpomtot .
We also show that the data are consistent with a constant ($s$-independent) $K$.

Longitudinal event structure and energy-flow measurements 
for React.~\ref{eq:dif} are given in Chapt.~\ref{longi}.
Pronounced ``beam jets'' are seen in this energy flow 
in the \pom\ hemisphere, where there is almost complete acceptance.

\section{Apparatus}
\label{appar}
\indent 

A detailed description of the UA8 apparatus, its properties, triggering
capabilities and interface to the UA2 experiment are given 
elsewhere\cite{ua8hard}. Thus, we only provide here a brief summary
of the spectrometer.
Since UA8 was installed in the same interaction region
as the UA2 experiment\cite{ua2hard} and a common data acquisition system
was used\cite{ua8hard},
the data from the UA2 calorimeter system could 
be used offline to study the $X$ system in React.~\ref{eq:dif}.

\subsection{Roman-pot spectrometers}
\indent 

\begin{figure}
\begin{center}
\mbox{\epsfig{file=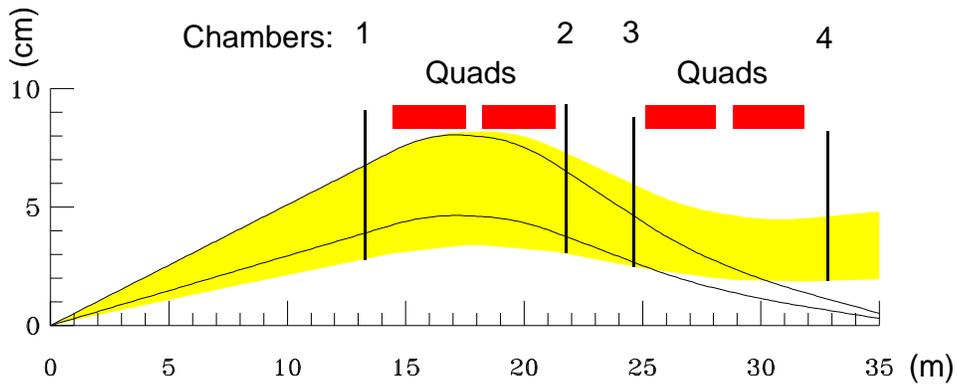,width=12.7cm}}
\end{center}
\caption[]{
Particle trajectories through a Roman-pot spectrometer. 
The labels, ``Quads", refer to the low-$\beta$ machine quadrupole magnets.
The center of the UA2 detector is at $z = 0$ at the left side of the sketch. 
The vertical lines indicate the positions
of the UA8 wire chambers in the Roman-pots.
The solid curves
show the trajectories of 300 GeV particles ($\xp \sim 0.95$)
emerging from the center of the
intersection region with minimum and maximum acceptable angles. 
The lower (upper) edge of the shaded area corresponds to the
minimum (maximum) angle of an elastic track which is accepted.
The trajectory corresponding to the lower edge 
of the shaded region in this sketch is 
12 beam widths ($\sigma$) from the center of the circulating beam orbit.
}
\label{fig:ua8traj}
\end{figure}

The Roman-pot spectrometers, which used the low-$\beta$ machine quadrupole
magnets, consisted of four pot installations positioned 
in each arm of intersection LSS4 at the CERN \SPS -Collider.
The positions of the pots in one spectrometer are shown in 
Fig.~\ref{fig:ua8traj}. 
The four spectrometers are labled according to which arm they are in
(P for outgoing proton and M for outgoing antiproton),
and whether they are above or below the beam pipe (``U'' for ``Up'' and ``D'' 
for ``Down''). Thus, the ``Up'' spectrometer in the proton arm is called PU.
If a track is in a ``Down'' spectrometer, we define the ``adjacent" 
spectrometer to mean the ``Up'' spectrometer of the same arm. 
Similarly, we define ``opposite'' spectrometer to be the one 
diagonally opposite the one containing a trigger particle 
(i.e., in the other arm).

Figure~\ref{fig:ua8traj} shows inelastic ($\xp \sim 0.95$)
particle trajectories through one of the spectrometers.
The shaded region shows the allowed trajectories for elastic scattering.
The final state proton or antiproton momentum
is calculated using the reconstructed vertex position (if it exists),
given by the UA2 central chamber system 
and points reconstructed from hits in chambers
1, 2 and 3. Chamber 4 was also used in the fit, if a track traversed it.

The UA4 experiment\cite{ua4hard}, in their 
measurements of elastic and diffractive scattering\cite{ua4dif1,ua4dif2}, 
used similar pot installations in LSS4, but 
without a pot upstream of the first quadrupole magnets.
With this first measuring station, our installation yielded 
acceptance for leading protons and antiprotons
with \xp\ as small as 0.70.

A system
of wire chambers was equipped with high bandwidth readout electronics
which interfaced to a 240 MIPS (Million Instructions per Second) 
data-driven trigger processor\cite{ua8hard,ddp} 
for calculating the proton or antiproton momentum online.
High speed, efficient triggering was necessary because we were interfaced to 
the UA2 data acquisition system and were required to present
only minimum perturbation to that experiment.

As discussed in Ref.~\cite{ua8hard},
each chamber contains 6~planes with 2~mm wire spacing in a conventional
MWPC configuration. There are 2 planes with horizontal wires, $y$ and $y'$
(shifted half a cell), and two sets of $\pm 7^\circ$ stereo 
views (with respect to the horizontal), $u$, 
$u'$ and $v$, $v'$. With a 4-bit time-to-digital converter (5~ns least count)
on each wire, a chamber provides $\sim 65 \, \mu$m position resolution
in the vertical (bending) plane.

\begin{figure}
\begin{center}
\mbox{\epsfig{file=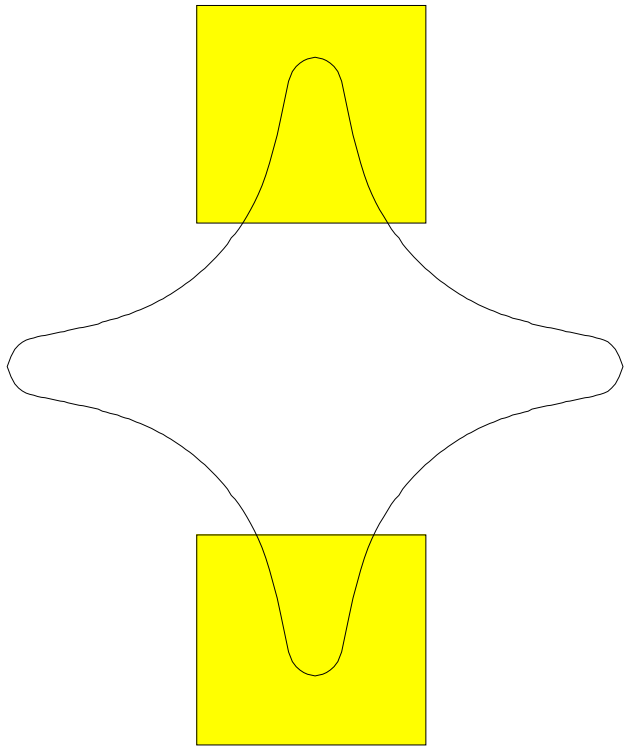,width=12.7cm}}
\end{center}
\caption[]{
The UA8 spectrometer aperture viewed from the interaction region. 
The shaded rectangles indicate the sensitive regions of the first wire chambers
at a distance $z$ = 13 m from the interaction region center.
The curved line indicates the walls of the beam pipe inside the quadrupole 
magnets.
}
\label{fig:aperture}
\end{figure}

Figure~\ref{fig:aperture} shows a ``beams-eye'' view of the UA8 chamber
aperture which is closest to the center of the interaction region.
The four-lobed curve in the figure illustrates the contour of the 
beam pipe which follows that of the quadrupole-magnet pole pieces.
The overlap between the beam pipe and rectangular chambers above and below
the beam illustrates the limited azimuthal range through which a
particle may pass.
A discussion of the acceptance corrections for the resulting losses 
is given below in Sect.~\ref{geomacc}.
Data were recorded with the bottom edge of each pot set, in different runs,
at either 12 or 14 beam widths from the beam axis. The combined data sample
is referred to below as the ``12$\sigma$ and 14$\sigma$'' data.

The Roman-pot spectrometers in the $\bar{p}$-arm were also used in a
stand-alone mode for inclusive measurements\cite{lambda,ua8hard} 
of $\bar{\Lambda}$ production for $x_{\Lambda} > 0.7$.

\subsection{Spectrometer resolutions}
\indent

\begin{figure}
\begin{center}
\mbox{\epsfig{file=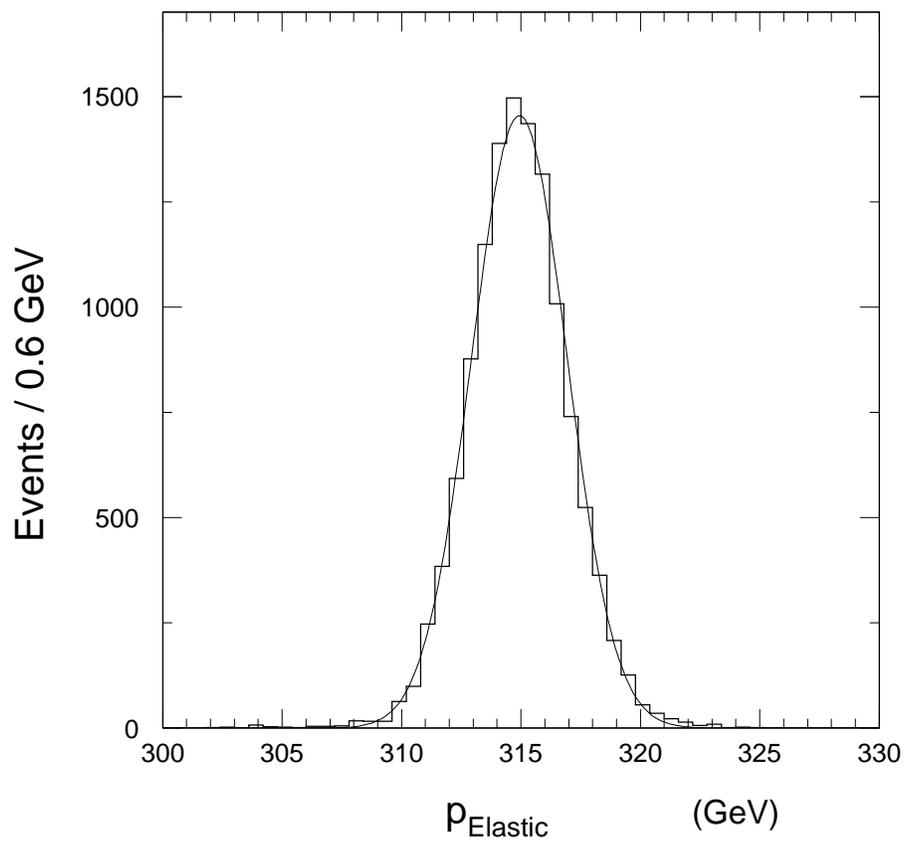,width=12.7cm}}
\end{center}
\caption[]{
An elastic momentum spectrum measured by UA8. 
The curve is a Gaussian fit
to the data histogram and has a width, $\sigma _p / p$ = 0.0061.
}
\label{fig:elastic}
\end{figure}

Figure~\ref{fig:elastic} shows the momentum distribution of tracks in elastic
scattering events (events with
two collinear tracks, no evidence of other tracks in the 
event, and no energy in the UA2 calorimeter system) 
with a Gaussian curve fitted to the data. 
The fitted $\sigma$ = 1.95 GeV, or $\sigma (\xp )$ = 0.0061, 
implies a resolution in diffractive mass,
of $\sigma(\rsp )$ = 1230/\rsp\ GeV.
This resolution improves with increasing scattering angle,
approximately as 0.0077/$|t|$,
due to the characteristic field shape of the quadrupole magnets. 

The resolution in momentum transfer, \T , is dominated by the dispersion 
of the incident beams resulting from the low-$\beta$ injection. The beam 
dispersions have been calculated to be 180 $\mu$rad and 150 $\mu$rad in the 
vertical planes of the $p$ and \ap\ beams, respectively. Since the
scattering angle in the vertical plane, 3--5 mrad, 
is much greater (for accepted particles) than the
average scattering angle in the horizontal plane, 0.2 mrad, the 
vertical component dominates the uncertainty. This leads to a \peetee\
uncertainty of about 0.052 GeV, and \T\ uncertainties from 
0.10 to 0.15 GeV$^2$ in the range, $1.0 < -t < 2.2$~GeV$^2$.

An additional \T -scale uncertainty exists due to the uncertainty in the 
absolute vertical
position of the Roman-pots relative to the beam line. This position
uncertainty is estimated to be less than 300 $\mu$m, which corresponds to a 
23 $\mu$rad shift in the average measured angle
or to \T -scale shifts of 0.014 to 
0.021 GeV$^2$ in the range, $1.0 < -t < 2.2$~GeV$^2$. 
From all \T\ uncertainies, there is a combined 
systematic cross section uncertainty of about 6\%.

\subsection{UA2 calorimeter and time-of-flight counters}
\label{ua2}
\indent

\begin{figure}
\begin{center}
\mbox{\epsfig{file=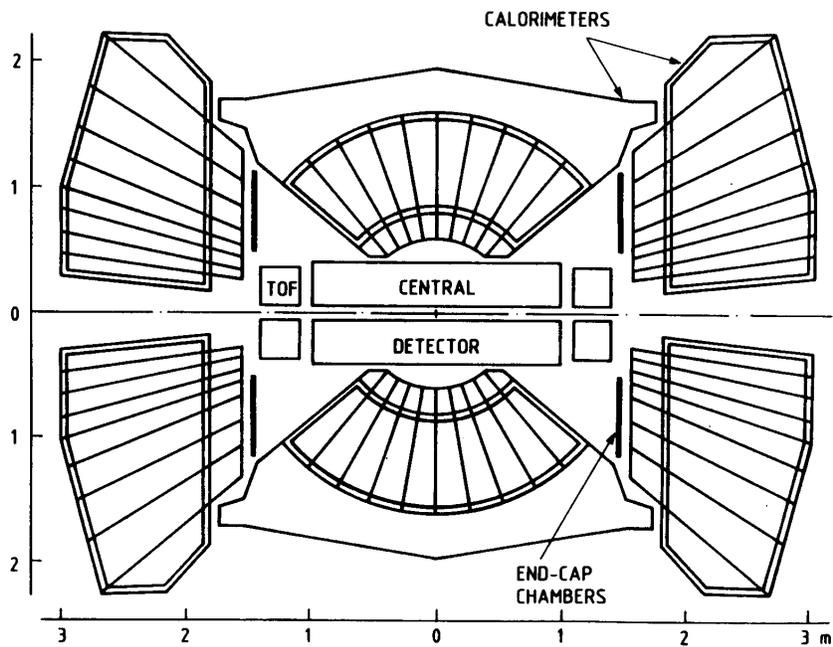,width=12.7cm}}
\end{center}
\caption[]{
A cross sectional view of the upgraded UA2 apparatus. 
Detectors which were important for the 
analysis described here are the Calorimeters, the Time-of-Flight counters
(TOF) and the Silicon Vertex Detector within the Central Detector assembly.
The TOF counters covered the pseudorapidity range, 2.3 to 4.1 in each arm.
}
\label{fig:ua2}
\end{figure}

Figure~\ref{fig:ua2} shows the UA2 calorimeter system
after the upgrade\cite{ua2hard}, in which end-cap calorimeters were added
to increase the acceptance in laboratory polar angle to
$6^\circ < \theta < 174^\circ$. The central calorimeter covers the
full azimuthal range, with twenty-four $15^\circ$ cells in $\phi$, and 
$140^\circ > \theta > 40^\circ$ with ten $10^\circ$ 
cells in $\theta$. The
end-cap calorimeters each adds 8 cells in $\theta$ and also covers
the full range of azimuthal angle. A tower structure is used which
points to the center of the interaction region.

Figure~\ref{fig:ua2} also shows the UA2 time-of-flight (TOF) counters, 
which cover the ranges $2^{\circ}$--$12^\circ$ and 
$168^\circ$--$178^\circ$ (with pseudorapidity,  $2.3 < |\eta| < 4.1$).
As discussed below in Sect.~\ref{topo}, when used offline together with the
UA2 calorimeter information, these counters offered us 
a means to define various classes of event topologies.
For example, these counters were found to be very useful as offline vetoes,
for defining pseudorapidity gaps in both arms as a signature of 
double-\pom -exchange\cite{pompom}.
 
The use of the updated version of the UA2 detector simulation
software\cite{ua2mc}, with which we could simulate observed data
for any assumed physics model, was very important in all
analyses involving data from the UA2 detectors. 

\subsection{Data acquisition trigger}
\label{daqtrig}
\indent

The trigger logic used to record examples of React.~\ref{eq:dif} contained 
the following components:
\begin{itemize}

\item
The wire chamber data were used in the data-driven trigger processor, as
described in the following paragraphs, to determine that a valid track existed
in at least one spectrometer.

\item
The scintillation counters were used to veto events that had: 
(a) beam halo tracks, or (b) hits
in a spectrometer adjacent (i.e. in the same arm) to one with a valid track.
\end{itemize}

The data-driven trigger processor\cite{ua8hard,ddp} 
required that there be either one hit or two contiguous
hits in the $y$ and $y'$ planes in each of the first three chambers 
of a spectrometer.
Moreover, the $y$ and $y'$ hits in each plane were required to have 
sufficient proximity to be consistent with the passage of a single track 
through the chamber. If these conditions were satisfied, the processor 
calculated the momentum 
from the tracks passage through the quadrupole magnet(s). 

\begin{figure}
\begin{center}
\mbox{\epsfig{file=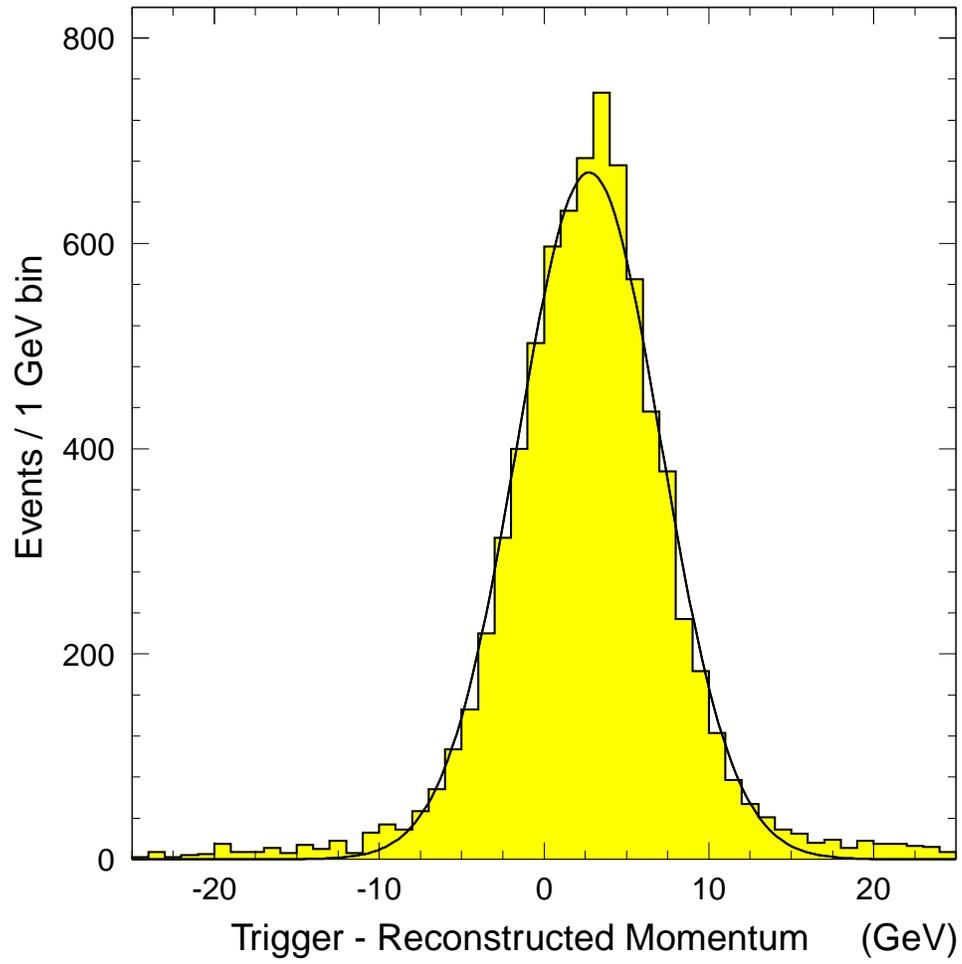,width=12.7cm}}
\end{center}
\caption[]{
The difference between momentum calculations made
by the trigger processor and in the offline analysis. The curve is a
Gaussian fit with $\sigma$ = 4.4 GeV. The offset of the mean value 
from zero is discussed in the text.
}
\label{fig:dimresol}
\end{figure}

The online momentum resolution of the trigger processor is illustrated in 
Fig.~\ref{fig:dimresol} which shows the difference between the offline
and online momentum calculations for a sample of proton trigger events. 
The Gaussian fit
with $\sigma = 4.4$ GeV corresponds to a processor momentum resolution of 
1.4\%. The fact that this resolution is worse than the offline
resolution of 0.6\% stems from the facts that the 
processor momentum estimate is based on second-order transport equations
through the magnets and that the primary vertex point was not used in the 
online algorithm.
Furthermore, the offline estimate benefits
from improved chamber alignment and drift time corrections.  
However, there is no reason why an improved online algorithm could not
be used in any future experiment of this type.
The distribution  in
Fig.~\ref{fig:dimresol} does not peak at zero because of small errors in
the chamber positions which were used to compute the 
tables used online in the processor. 

The trigger was set to select tracks with \xp\ in the range 0.70 to 1.05.
The trigger processor decreased the trigger rate 
by a factor of $\approx 150$ from the raw scintillator coincidence rate,
with no significant loss of good proton events. 

For each spectrometer, scintillation counter
signals were used to veto events with beam halo tracks
by demanding that there be no counter hit in
planes 1 or 2 of the diagonally opposite spectrometer, 
which had the timing of an incoming beam bunch.

The data recorded for React.~\ref{eq:dif} 
relied entirely on the Roman-pot spectrometers
and made no requirements on any of the UA2 detectors. 
However, the UA8 trigger was OR'ed with the normal UA2 triggers and all 
UA8 and UA2 detector data were written out to UA2 data tapes for each UA8 
trigger. Thus, in the offline analysis, the UA2 detector information
was available for each example of React.~\ref{eq:dif}. 

\section{Event selection and corrections}
\label{select}
\subsection{Event selection and background sources}
\indent

Table~\ref{tab:select} shows how the event data sample\cite{ayse} 
is reduced with each
of the offline selection cuts. 
Of the initial event sample of 150,000 triggered events 
(combined 12$\sigma$ and 14$\sigma$), 
59\% of all triggers are found with one reconstructed track 
which has $\xp > 0.9$ (this fraction would have been much larger, had we
increased the lower limit of $\xp > 0.7$ imposed in the trigger processor). 
A small number of these events (1.7\%)
had hits in the adjacent spectrometer and were rejected. 
Since this cut was in the hardware trigger, the offline cut serves 
to remove veto inefficiencies.

10.5\% of these events contain evidence that
two or more interactions occurred in the same bunch crossing (``pileup'')
and are removed from the event 
sample\footnote{In the rejected events, a measured $p$ or \ap\ from a 
bona fide 
example of React.~\ref{eq:dif} has associated calorimeter information which is 
augmented by particles from a different event.}. 
In the accepted events,
the UA2 and UA8 time-of-flight
information is required to be consistent with the single event hypothesis,
at most one vertex is found using the UA2 silicon detectors, 
and the total visible longitudinal momentum (calorimeter + spectrometer) 
in the trigger hemisphere is less than the maximum possible beam momentum
(taking resolution into account).

A further cut removed 14.0\% of the remaining events because of contamination 
from beam halo tracks. 
This cut, which is described in more detail in Ref.~\cite{ua8hard}, is
based on the timing information of hits in the chambers of the diagonally
opposite spectrometer,  which is characteristic of a halo track passing through
the entire spectrometer system. 
Other evidence for halo contamination can be obtained from observation
of the transverse position of an event vertex (or from the 
transverse positions of the tracks at z = 0, if no vertex is found).
The final cut, called ``Energy Topology'' in Table~\ref{tab:select}, 
is described in the next section.

\begin{figure}
\begin{center}
\mbox{\epsfig{file=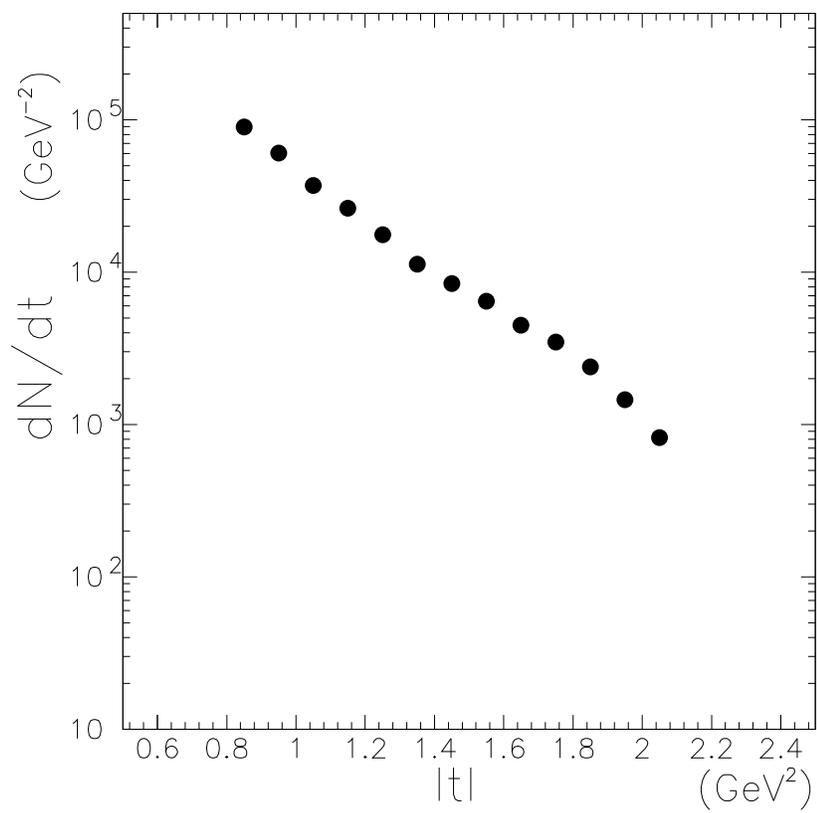,width=12.7cm}}
\end{center}
\caption[]{
Raw (uncorrected) momentum transfer (\T ) distribution 
of events with an antiproton in the upper \ap\ spectrometer.
The plot contains both 14$\sigma$ and 12$\sigma$ data.
}
\label{fig:rawt}
\end{figure}

\begin{figure}
\begin{center}
\mbox{\epsfig{file=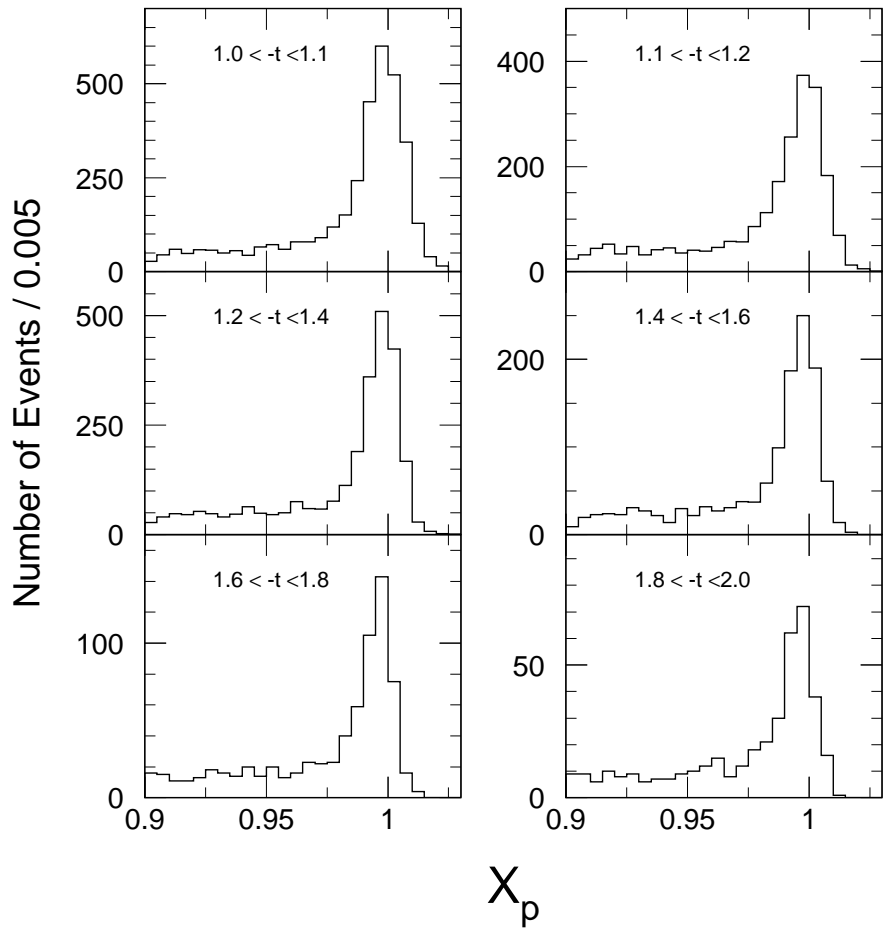,width=14cm}}
\end{center}
\caption[]{
Uncorrected $\xp = 1 - \xi$ distributions in the indicated \T -bins,
where $t$ is given in units of GeV$^2$.
}
\label{fig:dndx}
\end{figure}

Figure~\ref{fig:rawt} displays the uncorrected distribution in momentum
transfer, \T , for the final sample of 63K events with $\xp > 0.9$.
The distribution reflects the smooth acceptance function defined
below.  
Figure~\ref{fig:dndx} shows the corresponding uncorrected 
\xp\ distributions for several ranges of \T .
There is seen to be a small improvement in the resolution in the high $|t|$ 
region due to the quadrupole field shape. 

\begin{figure}
\begin{center}
\mbox{\epsfig{file=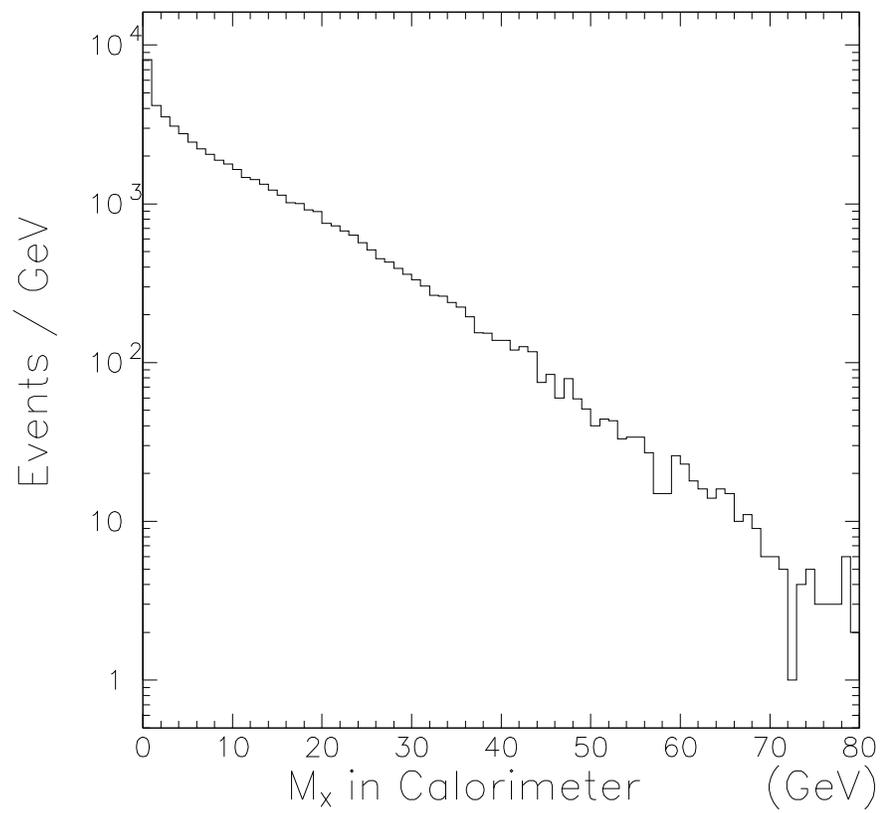,width=12.7cm}}
\end{center}
\caption[]{
The uncorrected total invariant mass seen by the UA2 calorimeter 
for events with $\xp > 0.9$.}
\label{fig:calmas}
\end{figure}

Figure~\ref{fig:calmas} shows the invariant 
mass of the portion of the $X$ system  
in React.~\ref{eq:dif} which was contained in the UA2 
calorimeter\footnote{The calorimeter
invariant mass is estimated for each event by assuming
that non-zero energy in each ``struck" cell of the calorimeter
is caused by a massless particle and then calculating,
$M_X^2 = (\Sigma E_i)^2 - |\Sigma \vec{P}_i|^2$, summing over all cells.}.
Due to incomplete polar angle coverage of the calorimeter system, 
particles can completely miss the calorimeter at small angles, resulting
in a calculated invariant mass which is less than $\rsp$. 
In particular, the peak at zero mass in Fig.~\ref{fig:calmas}
is due to low-mass diffractive events whose particles
in the diffractive system are all sharply collimated forward, 
opposite the observed proton.

\subsection{Use of energy-flow topology in event selection}
\label{topo}
\indent

The availability of the 
data from the UA2 calorimeter and time-of-flight (TOF) counters
(pseudorapidity range $2.3 < \eta < 4.1$)
in the offline analysis
allows us to study the energy-flow topology of the diffractive system.
Such information permits the removal of certain types of background,
such as when the measured diffractive mass, \rsp , is inconsistent 
with the observed energy-flow topology. This can occur, for example,
if the observed $p$ or \ap\ is not alone at the lower vertex in 
Fig.~\ref{fig:difdiag}.
 
\begin{figure}
\begin{center}
\mbox{\epsfig{file=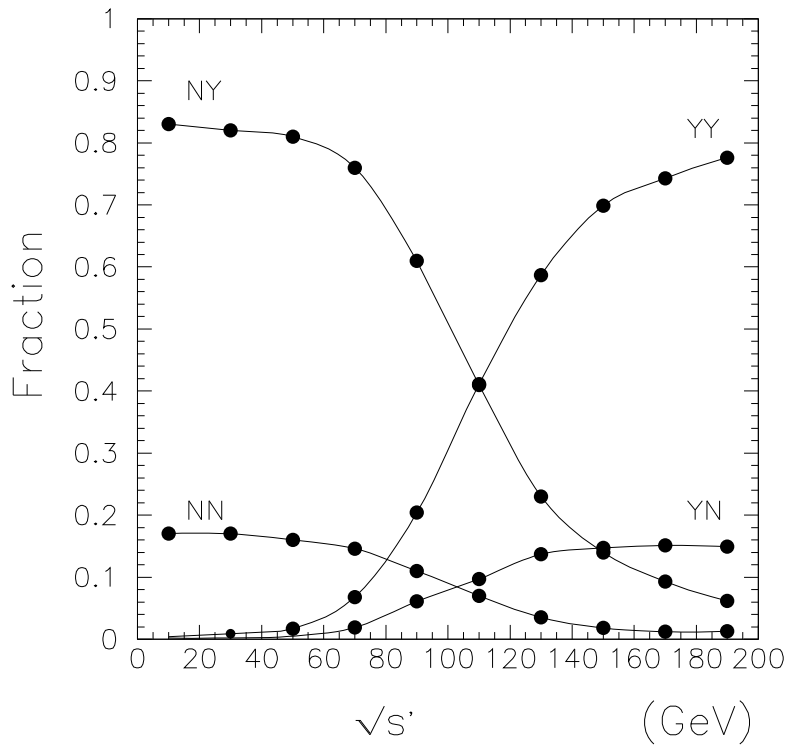,width=12.7cm}}
\end{center}
\caption[]{
Dependence on diffractive mass of the four different 
TOF-counter hit topologies.
The curves show the relative frequencies of 
events with no TOF counter hits (NN), 
hits only opposite the detected $p$ or \ap\ (NY), 
hits only on the same side as the detected $p$ or \ap\ (YN) 
and hits in both sides (YY), respectively.
The experimental mass resolution, $\sigma(\rsp )$ = 1230/\rsp\ GeV,
results in some ``smearing" of the NY and NN curves at low mass.
}
\label{fig:toftop}
\end{figure}

Figure~\ref{fig:toftop} shows the fractions of all events which have
hits in the four combinations of trigger-side and opposite-side TOF counters,
plotted vs. diffractive mass, \rsp . 
These four possible combinations
of struck TOF counters are as follows:
\begin{itemize}
\item NY: (``No-Yes'') 
Trigger-side counters not hit; opposite-side counters hit.
\item YN: Trigger-side counters hit; opposite-side counters not hit. 
\item YY: Both trigger-side and opposite-side counters hit. 
\item NN: No TOF counters hit.
\end{itemize}

The dominant patterns in Fig.~\ref{fig:toftop} are seen to be NY and YY.
As expected, NY dominates for low mass diffraction;
NN is a smaller
subset of the physics of NY, when a low-mass diffractive system
is forward at such a small angle that its tracks miss the TOF counters.
YY dominates at high mass,
corresponding to a forward cone of particles which
increasingly spreads out into the trigger hemisphere at larger
mass. In addition, however, there is the smaller component, YN, 
in which only  counters on the trigger side are hit.

In order to further understand the topologies of 
these events, we use the calorimeter
information and define the variable, $\rm \theta_{cal}$, which for each 
event is the 
polar angle in the laboratory of the vector sum of its calorimeter
energy vectors (the energy in each cell is treated as a vector). 

\begin{figure}
\begin{center}
\mbox{\epsfig{file=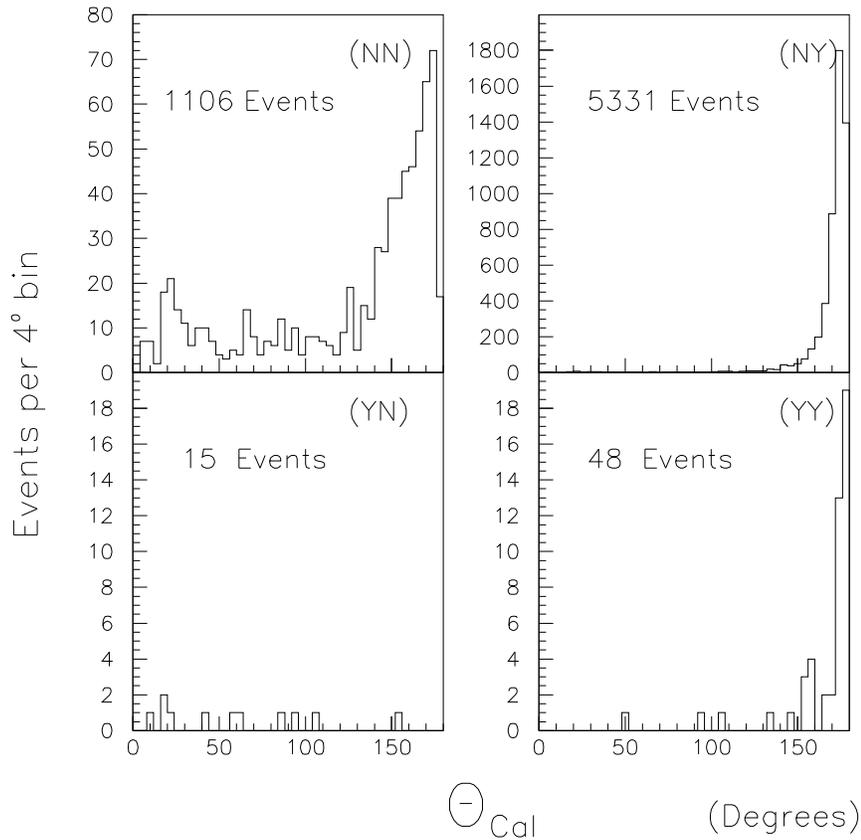,width=12.7cm}}
\end{center}
\caption[]{
Angular distributions of the calorimeter-energy-sum vector for
the four TOF topologies, selected for low-mass at the center
of the diffractive peak ($\xp \sim 1$). 
As in Fig.~\protect\ref{fig:toftop}, 
the labels refer to the TOF counters which have hits 
(e.g., NY means ``No'' for counters on the same side as the detected proton
and ``Yes'' for counters on the opposite side). 
Note that, in the NN (NY) plots, 443 (230) of
the 1106 (5331) events do not appear in the histogram, because there is no
energy deposited in the calorimeter.
The trigger side is at 0$^{\circ}$.
}
\label{fig:tcal30}
\end{figure}

\begin{figure}
\begin{center}
\mbox{\epsfig{file=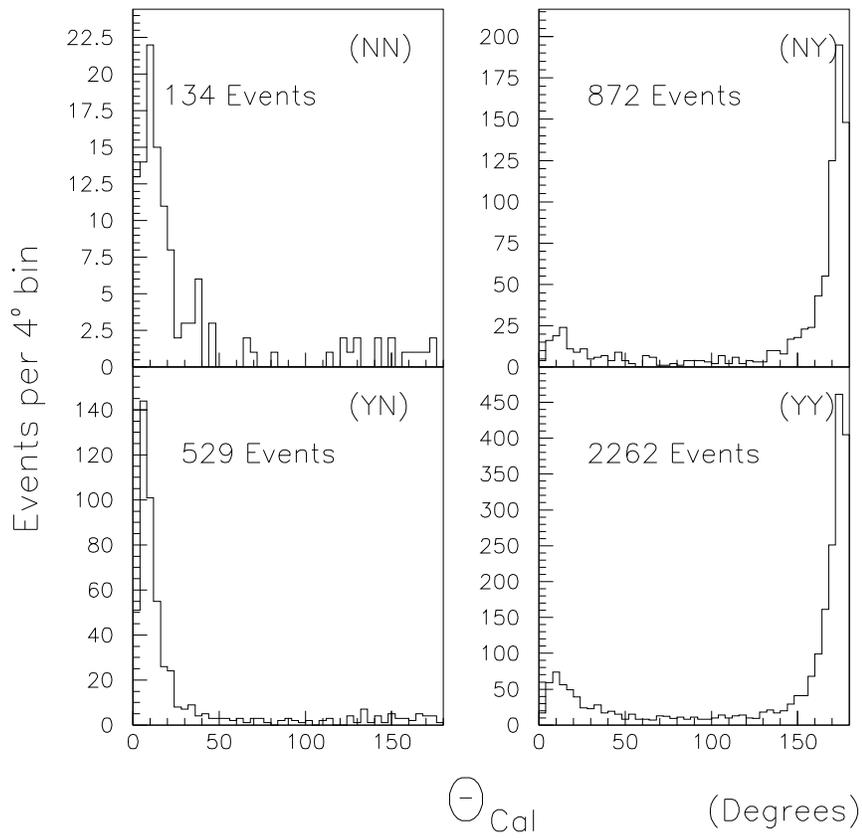,width=12.7cm}}
\end{center}
\caption[]{
Angular distributions of the calorimeter-energy-sum vector for
the four TOF topologies in diffractive events with \protect\rsp\ = 130 GeV.
As in Fig.~\protect\ref{fig:toftop}, 
the labels refer to the TOF counters which have hits 
(e.g., NY means ``No'' for counters on the same side as the detected proton
and ``Yes'' for counters on the opposite side).
The trigger side is at 0$^{\circ}$.
}
\label{fig:tcal130}
\end{figure}

Figures~\ref{fig:tcal30} and \ref{fig:tcal130} show histograms 
of $\rm \theta_{cal}$ for each of the TOF patterns, for two event samples, 
the first (low-mass) selected from the center of the diffractive peak
at $\xp \sim 1$ and the second with \rsp\ = 130 GeV (\xp\ = 0.043).
At low mass, the NY event sample is dominant with a 
clear forward peak in $\rm \theta_{cal}$ (hits opposite the trigger proton).
A forward peak also exists at low mass in the NN sample, corresponding
to events in which high momentum tracks in the diffractive system
have $\eta > 4.1$, some low momentum tracks hit the calorimeter cells
at small angles ($\eta < 2.3$), but the opposite-side TOF counters are 
not hit.
A very small and inconsequential signal exists in the YY class, where the
trigger proton is accompanied by one or more low energy tracks which hit
the same-side TOF counters. 

In the higher mass sample with \rsp\ = 130 GeV in Fig.~\ref{fig:tcal130}, 
the YY and NY topologies
dominate with clear peaks in the $\rm \theta_{cal}$ distributions opposite
the trigger particle. As expected, this is due to a spreading of lower momentum 
particles from the diffractive system into the trigger-side hemisphere at
higher mass. 
However, a new class of events occurs in Fig.~\ref{fig:tcal130}.
In the YN and NN samples, where the TOF counters in the opposite arm have no 
hits, $\rm \theta_{cal}$ is peaked on the trigger side. 
We identify this effect
with diffractive excitation on the trigger side.
As such, these events constitute a class of background events and are
removed from our data sample in 
Table~\ref{tab:select} if $\rm \theta_{cal}< 90 ^\circ$.

\begin{figure}
\begin{center}
\mbox{\epsfig{file=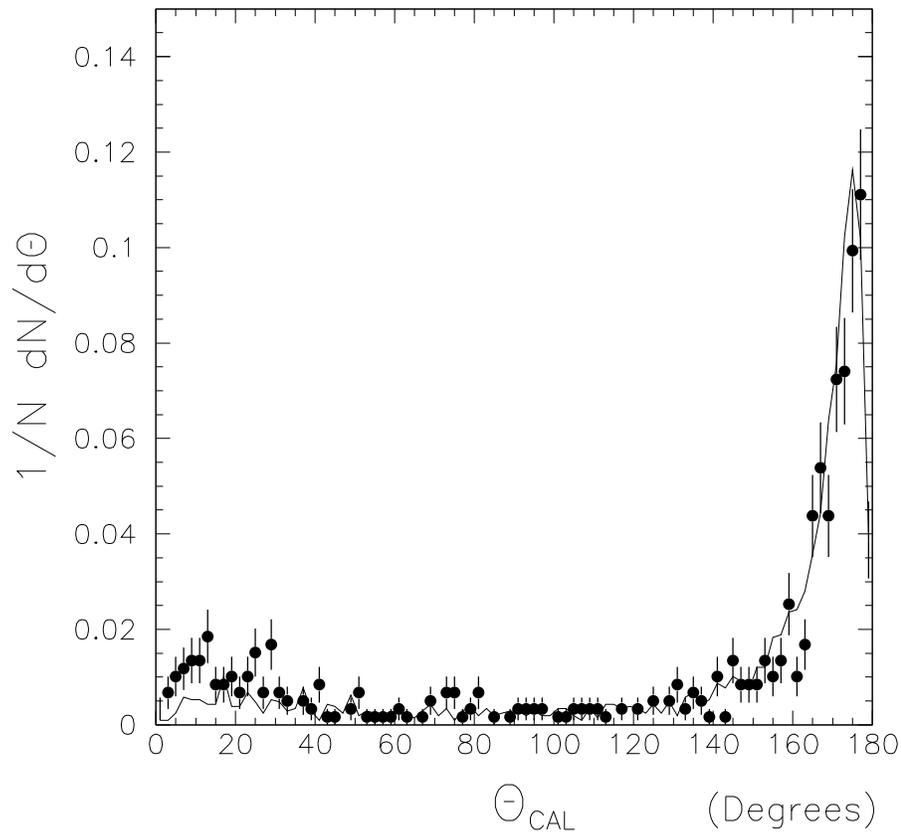,width=12.7cm}}
\end{center}
\caption[]{
Polar angle of the energy sum vector in the UA2 calorimeter system, 
$\rm \theta _{cal}$, 
for NY and YY events with \rsp\ = 130 GeV  
(see Fig.~\protect\ref{fig:tcal130}) and total transverse 
calorimeter energy in the range, $5 < \protect\sumet < 10$~GeV.
The curve from the Monte Carlo calculation is normalized to the data.
}
\label{fig:tcalmc}
\end{figure}

We can compare the dominant NY and YY events with a Monte Carlo 
model\cite{p4pi} of longitudinal (or $p_t$-limited) diffractive excitation
(see Section~\ref{longi}).
Figure~\ref{fig:tcalmc} shows $\rm \theta_{cal}$ for NY and YY 
events at \rsp\ = 130 GeV, which have total transverse 
calorimeter energy in the range, $\rm 5 < \Sigma E_{t} < 10$ GeV. 
Good agreement is seen between data and Monte Carlo, except at small 
$\rm \theta_{cal}$,
where there appears to be a small excess of events over the Monte Carlo
calculation. This could be due to a residual background from
double-diffractive events, where the opposite side diffractive system 
strikes the TOF counters.

\subsection{Efficiencies for trigger and selection cuts}
\indent

Table~\ref{tab:efficac} 
lists the individual efficiencies for the trigger components
discussed in Sect.~\ref{daqtrig}, as well as for 
the pileup and halo offline cuts listed in Table~\ref{tab:select}.

The two trigger processor corrections were determined by the
combined use of processor emulation in the UA8 detector simulation
Monte Carlo program and by offline analysis of data recorded without
using the processor in the trigger. The two small corrections for
accidental vetoes were estimated from the measured halo and noise rates,
respectively.

\begin{figure}
\begin{center}
\mbox{\epsfig{file=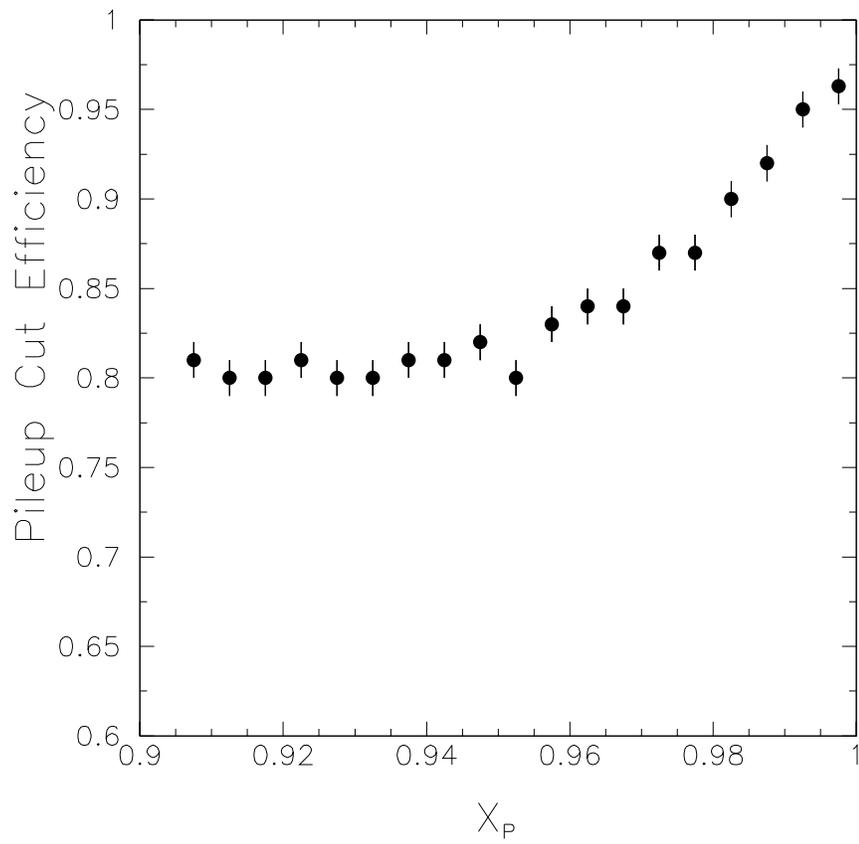,width=12.7cm}}
\end{center}
\caption[]{
The fraction of good events which survive the pileup cut 
(see description in text) as a function of the observed track's \xp .
}
\label{fig:xpileup}
\end{figure}

Events with a good spectrometer track (i.e., not beam halo), 
which are removed from the sample by the offline pileup 
selection described in Sect.~\ref{select}, 
are corrected for in the cross section evaluation.
This is accomplished by determining how many events in the pileup-rejected 
event sample have good spectrometer tracks. 
Figure~\ref{fig:xpileup} shows the fraction of events in the total sample which
survive the pileup cut as a function of track \xp.
Although the 
average loss due to pileup is 9.2\%, we use the distribution shown in
Fig.~\ref{fig:xpileup} to obtain the final cross section numbers given below.

The fraction of good events that are accidentally rejected by the 
offline halo veto is given by the number of halo-rejected
events with a single vertex (as determined using the UA2 silicon detectors), 
which pass the pileup cut, divided by the total
number of single-vertex events. This fraction is 8.4\%.

\subsection{Geometric acceptance corrections}
\label{geomacc}
\indent

Monte Carlo detector simulation software was written to
correct for all geometric acceptance and detector inefficiencies.
It was also used to estimate resolution effects whose origins are related to 
effects such as: 
$\rm t_o$-jitter in the chamber TDC system, time slewing, digitization bins 
and trigger requirements. 
The simulated data were passed through the 
normal pattern-recognition software for 
decoding and analysis, with the identical procedures 
used in the processing of real data. The resulting resolutions in
chamber spatial measurements, elastic momentum and other variables
are in good agreement with those observed in the real data.

\begin{figure}
\begin{center}
\mbox{\epsfig{file=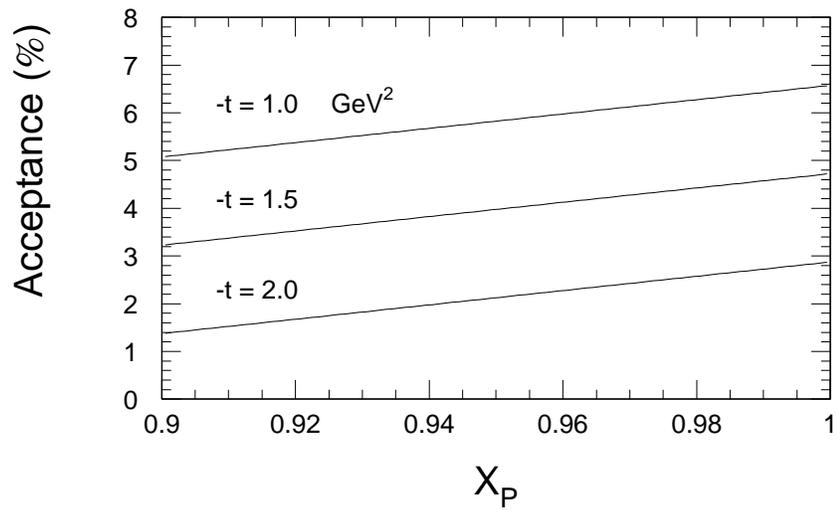,width=12.7cm}}
\end{center}
\caption[]{
Geometric acceptance of one spectrometer 
(upper \ap ) vs. \xp\ at three values of momentum transfer.
This acceptance corrects a single-spectrometer cross section for
full $\phi$-dependence in its arm.
See discussion in text.
}
\label{fig:muacc}
\end{figure}

Figure~\ref{fig:muacc} shows the geometric acceptance at
several \T -values plotted vs. \xp\ for the MU spectrometer.
In this figure, the geometric acceptance is defined as 
the fraction of all antiprotons produced in the antiproton arm
with a specific \xp\ and \T\ which is detected and reconstructed by
the MU spectrometer.
These acceptances are found to vary smoothly 
over the \xp\ and \T\ ranges covered in this paper.

\section{Absolute cross sections}
\label{sigmas}

\subsection{Differential cross section vs. {\boldmath \xp\ and \T }} 
\indent

The inclusive single diffractive differential cross section, \dsig ,
has been evaluated in the \xp\ range, 0.91--0.97, 
where the upper limit is chosen to avoid 
resolution ``smearing" from the peak at $\xp \approx 1$. 
This point is demonstrated by the dashed curve in Fig.~\ref{fig:mumom}
which is approximately a mirror image of the event distribution
with $\xp > 1.0$, the existence of which is a pure resolution effect.

\dsig\ was evaluated at points in $\xi = 1-\xp$ and \T\ 
space, independently for each of the
three spectrometers, PU, PD and MU (the fourth spectrometer was not used 
for the analysis reported here).
The resulting weighted averages at each point
are given in Table~\ref{tab:sigmatm}.

The average-\T\ values calculated for the events in each \T -bin show
no systematic shifts as a function of \xp\ and the average values
are close enough to the bin center values that the bin centers are used in
the analyses described below.

\begin{figure}
\begin{center}
\mbox{\epsfig{file=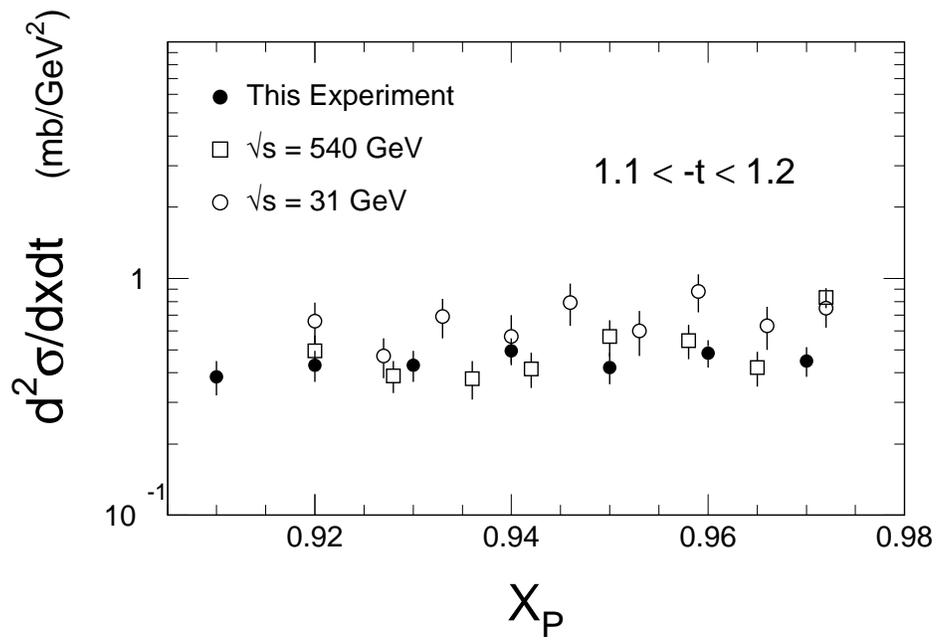,width=14cm}}
\end{center}
\caption[]{
UA8 differential cross section vs. \xp , for the $|t|$-bin, 1.1--1.2 GeV$^2$,
compared with results from Experiment UA4\protect\cite{ua4dif1} and the 
ISR\protect\cite{albrow}.
}
\label{fig:compisr}
\end{figure}

Figure~\ref{fig:compisr} compares our \dsig\ values in the \T -bin, 
1.1--1.2~GeV$^2$ with measurements in the same range at the 
\SPS -Collider\cite{ua4dif1} (UA4 experiment)
and at the CERN Intersecting Storage Rings\cite{albrow}.
The UA4 points are in reasonable agreement with our UA8 points, while the
ISR cross sections are somewhat larger, on the average. As will be shown in 
Sect.~\ref{sec:sdep}, the $s$-dependence of \dsig\ at fixed $\xi = s'/s$ and \T\
directly exhibits the $s'$-dependence of the \pom -proton total cross section,
wherever non-\pom -exchange background can be ignored. It is seen that
the \pom -proton total cross
sections at $\sqrt{s'} = \sim$6~Gev and $\sim$100 GeV are the same within
about $\sim 20\%$.

In our analyses, we make use of published 
\dsig\ data from the following experiments:
\begin{itemize}  
\item FNAL fixed target: Schamberger et al.\cite{schamberger},
$s$ = 262, 309, 366, 565, 741 GeV$^2$.
\item FNAL fixed target: Cool et al.\cite{cool}, $s$ = 189, 377 GeV$^2$.
\item FNAL fixed target: Akimov et al.\cite{akimov}, $s$ = 294, 700 GeV$^2$.
\item CERN ISR: Albrow et al. (CHLM Collaboration)\cite{albrow}, 
$s$ = 551, 930 GeV$^2$.
\item CERN \SPS -Collider: Bozzo et al.(UA4 Collaboration)\cite{ua4dif1}, 
$s$ = (546)$^2$ GeV$^2$.
\item FNAL Tevatron: Abe et al.(CDF Collaboration)\cite{cdf}, 
$s$ = (546)$^2$, (1800)$^2$ GeV$^2$ ({\it Fitted function only}).
\end{itemize}  

The main triple-Regge analysis reported below is done by simultaneously
fitting our UA8 data and the ISR data of Albrow et al.\cite{albrow}.
This procedure is dictated by the large amount of
tabulated data available in
Refs.~\cite{albrow} over a wide range of momentum transfer
and our wish to avoid using the lower energy data where resonance effects
distort the cross section (see Sect.~\ref{breakdown}). 

\subsection{Total single diffractive cross section, {\boldmath \sigdiftot}}
\label{totsig}
\indent 

Due to resolution smearing in \xp , it is not possible for us  
to quote differential cross sections for $\xp > 0.97$.
However, the cross section 
integrated over \xp , $d \sigma / dt$, can be evaluated because the acceptance
depends weakly on \xp\ in this region. 
It is conventional to quote $d \sigma / dt$
for $\xp > 0.95$, because the integral background from non-\pom\ exchange 
throughout this region can be ignored.

\begin{figure}
\begin{center}
\mbox{\epsfig{file=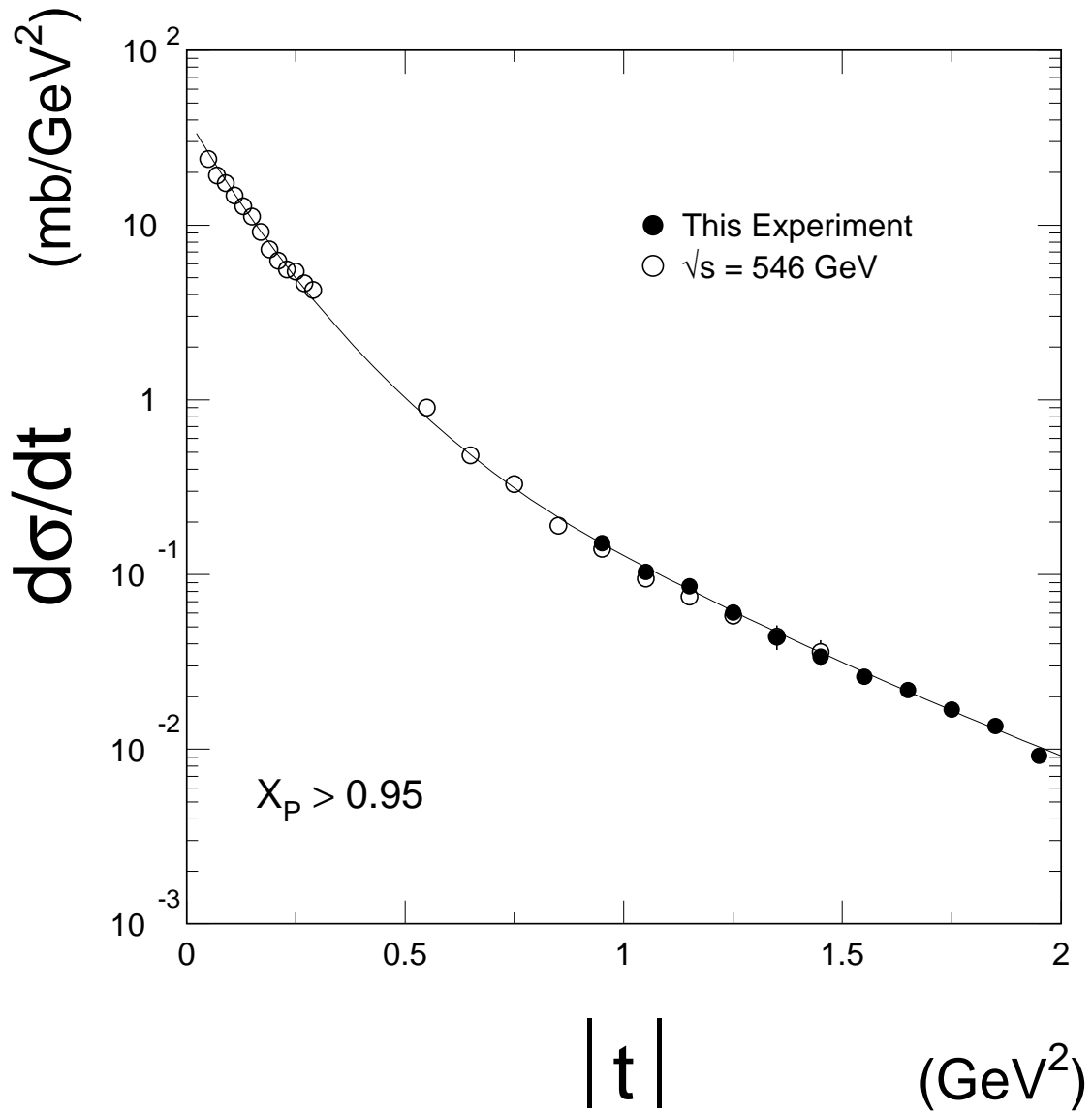,width=16cm}}
\end{center}
\caption[]{
Inclusive differential cross section for protons in React.~\protect\ref{eq:dif}
when $\xp > 0.95$, measured in this experiment and in experiment 
UA4\protect\cite{ua4dif1,ua4dif2} with \rs\ = 546 GeV. 
As in Table~\protect\ref{tab:dsigdt}, the cross sections shown are for
a single arm. 
Thus, the integral is one-half the total single-diffractive
cross section, \sigdiftot . The curve is to ``guide-the-eye".
}
\label{fig:dsigdt}
\end{figure}

The resulting (single arm) 
values of $d \sigma / dt$ are given in Table~\ref{tab:dsigdt}.
They are plotted in Fig.~\ref{fig:dsigdt}, together with
the corresponding measurements from the UA4 experiment\cite{ua4dif1,ua4dif2} 
at the
nearby energy of \rs = 546 GeV at the \SPS -Collider. The UA4 points
come from two independent runs, one at high-$\beta$ and one at low-$\beta$
which allowed them to span most of the available \T -range.  
The UA8 and UA4 points are seen to be in good agreement in the region where
they overlap. 
The solid curve superimposed on the points is to ``guide-the-eye".
UA4 
obtains\cite{ua4dif2} a total single diffractive cross section (both arms) of
($9.4 \pm 0.7$) mb, or $4.7 \pm 0.35$~mb for a single arm.

\begin{figure}
\begin{center}
\mbox{\epsfig{file=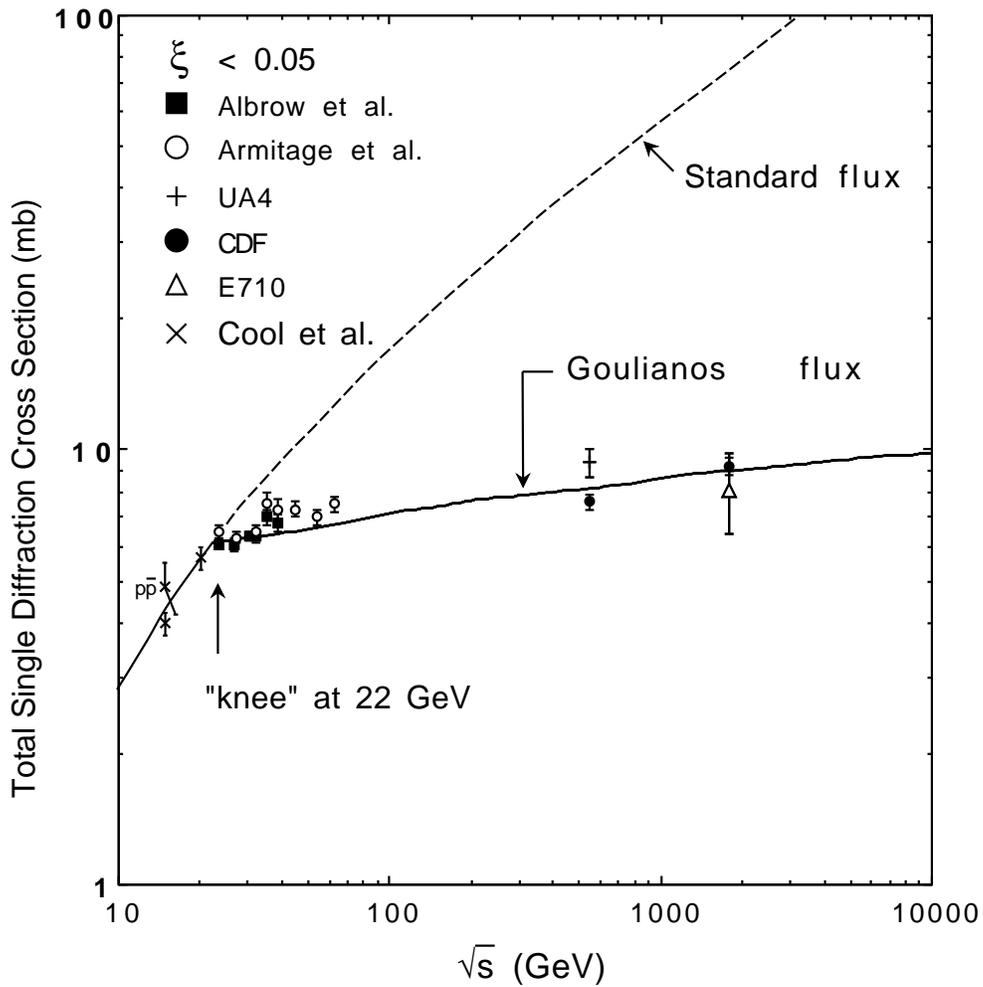,width=16.0cm}}
\end{center}
\caption[]{
Summary\protect\cite{dino} of existing measurements of the total single 
diffractive cross section (including a factor of 2x for both arms) for 
$\xi < 0.05$. Dashed curve is the prediction using the standard Triple-Regge
flux factor; solid curve is from Goulianos\protect\cite{dino}.
}
\label{fig:dino}
\end{figure}

Figure~\ref{fig:dino}, taken from Ref.~\cite{dino}, shows a 
survey of existing measurements of
\sigdiftot\ (for both arms) with $\xp > 0.95$, as a function of \rs . 
The points
at \rs\ = 13.8 and 19.4 GeV are the Fermilab fixed-target measurements of 
Cool et al.\cite{cool} at proton beam momenta of 100 and 200 GeV, respectively. 
A third Cool et al. 
point, marked $p\bar{p}$, is from data taken with a $\bar{p}$ beam.
At higher energy, there are the measurements at the
CERN Intersecting Storage Rings\cite{albrowtot,armitagetot} 
and, at the still higher energies of 
\rs\ = 546 and 1800 GeV, there are the measurements from the UA4, 
CDF\cite{cdf} and E710\cite{710} experiments.

Goulianos\cite{dino} has drawn attention to the remarkable
behavior of the total single diffractive cross section, \sigdiftot , 
as a function of energy.
The observed total single diffractive cross 
section does not continue to increase following the expected triple-Regge
behavior (see Sect.~\ref{phenom}), which is well known to violate unitarity.
The observed flattening of \sigdiftot\ for c.m. energies above about 20 GeV 
is associated\cite{dino} with a saturation of the flux factor, \flux .

The solid line in Fig.~\ref{fig:dino} represents an attempt\cite{dino} 
to describe the observed $s$-dependence of \sigdiftot . \sigpomtot\ is assumed 
throughout to follow the high-energy triple-\pom\ form: 
$\sigpomtot\ = \sigma_0 (s')^{\epsilon}$. Below \rs\ = 22 GeV, the 
normalization constant, $K$, in \flux\ is assumed to be constant, 
following conventional wisdom.
However, when $s$ exceeds $\sqrt{s} = 22$~GeV, $K$ is forced to decrease 
with increasing $s$, in order to account for the flattening of \sigdiftot .
This is accomplished by dividing $K$ by the 
integral over $\xi$ and \T\ of the Flux Factor.
In the following section, one of our principal concerns is to determine whether
the consequences of this prescription are in agreement with the measured
differential cross section, \dsig\ in React.~\ref{eq:dif}.

\section{Analysis}
\label{chaptanal}
\indent

In this section, we present a triple-Regge analysis of the
UA8 data together with the extensive measurements by the 
CHLM collaboration\cite{albrow} at the CERN Intersecting-Storage-Rings.
We explore the significance of the the fact that the 
differential cross sections
for React.~\ref{eq:dif} at the ISR and at the \SPS -Collider have similar
magnitudes.
We present a parametrization that describes all the data.

In the following sections, after summarizing the triple-Regge phenomenology 
which describes diffraction, we obtain values of the effective \pom\ 
Regge trajectory in the $|t|$--range, 1.0--1.6 GeV$^2$, which
shows a strong departure from linearity, 
with a flattening as $|\T|$ increases.

We then examine the $s$-dependence of \dsig\ at fixed $\xi$ and \T ,
which is seen to directly reflect the $s'$-dependence of \sigpomtot\
multiplied by the normalization of \flux , $K$. The data are seen to
have a behavior that is inconsistent with the $s$-dependent normalization
constant, $K(s)$, proposed by Goulianos\cite{dino}.
We also show that two components are required in \sigpomtot , corresponding
to the triple--\pom\ (${\cal P}{\cal P}{\cal P}$)
and \pom --\pom --Reggeon (${\cal P}{\cal P}{\cal R}$) terms
in the triple-Regge expansion.

Next, we present the results of 
successful fits to the combined data 
over the $\xi$ range, 0.03--0.09.
These fits are first done in the range, $\xi$ = 0.03--0.04, 
where the non-\pom -exchange background is small and therefore neglected, 
and then over the extended range, where the non-\pom -exchange
background is taken into account. The two types of fits are found to agree.
The requirement of a second term in \sigpomtot\ is  
reinforced by the fits. This is similar to what
is needed to describe real-particle total cross section data.
Independent confirmation of the nonlinear \pom\ trajectory is also obtained
from the fits.
We show that the non-\pom -exchange 
background in the differential cross sections cannot be accounted
for by one-pion-exchange.
Finally, we show that the triple-Regge formula is deficient in describing
data at low diffractive mass and energy (resonance region).

\subsection{Triple-Regge phenomenology of diffraction}
\label{phenom}
\indent

\begin{figure}
\begin{center}
\mbox{\epsfig{file=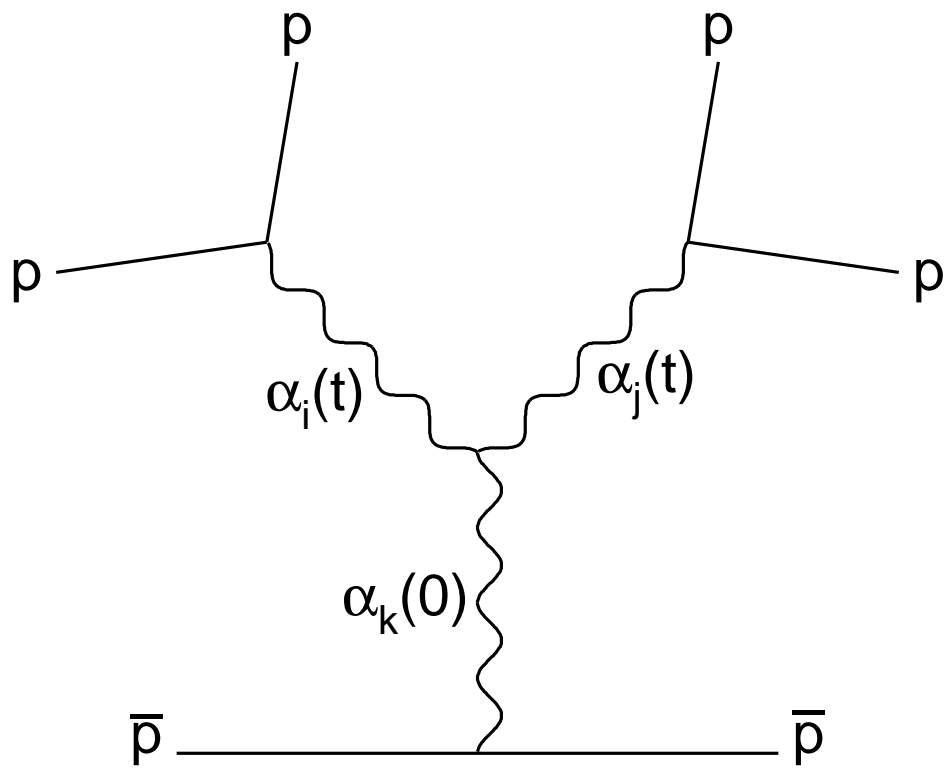,width=16.0cm}}
\end{center}
\caption[]{
Triple-Reggeon graph.
}
\label{fig:tripregg}
\end{figure}

Mueller's\cite{almueller} generalized optical theorem allows Regge  
analyses to be performed on inclusive processes such as React.~\ref{eq:dif}.
At small $\xi$ and $s' >> 1$, the Mueller approach yields a differential 
cross section of the form (see, for example, Ref.~\cite{fieldfox}):
\begin{equation}
{{d^2\sigma}\over{d\xi dt}}\, \, = 
\, \, \sum_{ijk} \, G_{ijk}(t) \x \xi^{1-\alpha_i(t) - \alpha_j(t)} 
\x (s')^{\alpha_k(0) - 1}
\label{eq:tp}
\end{equation}
$s'$ is in units of GeV$^2$.
$\alpha_i (t)$ is the Regge trajectory for
Reggeon $i$.
The sum is taken over all possible exchanged Reggeons in the
diagram shown in Fig.~\ref{fig:tripregg}. The $G_{ijk}(t)$ are
products of the various Reggeon-proton and triple-Reggeon couplings 
in Fig.~\ref{fig:tripregg} (and the signature factors). 
 
Because the \pom\ is the highest-lying trajectory (has intercept slightly
larger than unity at $t=0$), when $i=j=\pom $, $1 - 2 \alpha < 0$ and
the differential cross
section increases rapidly as $\xi \ra 0$. This corresponds to the empirical
observation that the most likely momentum fraction of the \pom\ in the
proton, $\xi$, is near zero.

There
are two dominant terms at small $\xi$, namely $ ijk = {\cal P}{\cal P}{\cal P}$
and ${\cal P}{\cal P}{\cal R}$. 
As discussed in Sect.~\ref{intro}, it has become customary to rewrite
Eq.~\ref{eq:tp} as a product of a \pom\ flux factor, \flux , in the proton 
(a measure of the probability to find a \pom\ in a proton with momentum
transfer, \T , and momentum fraction $\xi$)
and a \pom -proton total cross section, \sigpomtot .
The forms of \flux\ and \sigpomtot\ are as follows:
\begin{equation}
\label{eq:tripleR}
{{d^2 \sigdiftot}\over{d \xi dt}} 
\, = \,  \flux \x \sigpomtot (s') 
\, = \, [K \, |F_1(t)|^2  \, e^{bt} \, \xi^{1-2\alpha_{\cal P}(t)}] 
\, \x \, \sigma_0 [(s')^{\epsilon_1} \, + \, R \, (s')^{\epsilon_2}]
\end{equation}
where $|F_1(t)|^2$ is the standard Donnachie-Landshoff\cite{dl_elastic} 
form-factor\footnote{$F_1(t)={{4 m_p ^2 - 2.8t}
\over{4 m_p ^2 - t}}\, \x \, {1\over{(1-t/0.71)^2}}$} 
(see comment at the end of this section on the reason for the $e^{bt}$ term).
The squared diffractive mass, $s' = \xi s$.
The quantity, $\epsilon$, is defined as, $\alpha (0) = 1 + \epsilon$.

We note the following:
\begin{itemize} 
\item In Eq.~\ref{eq:tripleR}, $|F_1(t)|^2 \, \x \, e^{bt}$ 
carries the \T -dependence
of the $G_{ijk}$ in Eq.~\ref{eq:tp}. 
The same \T -dependence is assumed for both
$G_{{\cal P}{\cal P}{\cal P}}$ and $G_{{\cal P}{\cal P}{\cal R}}$.
Physically, this is saying that the \pom\ has the same flux factor in
the proton, independent of whether there is \pom -exchange or 
Reggeon-exchange in the \pom -proton interaction.
\item Expressing \sigpomtot\ as the sum of two components is in direct
analogy to all total cross sections, which are fit\cite{dl_tot,cudell,dino2}  
by the same 2-component Regge function:
\begin{equation}
\sigpomtot (s') \, \, = 
 \, \, \sigma_0 [(s')^{\epsilon _1} \, + \, R (s')^{\epsilon _2}]
\label{eq:twocomp}
\end{equation}
The first term corresponds to the triple-\pom\ process, 
(${\cal P}{\cal P}{\cal P}$),
while the second corresponds to other non-leading, C=+ ($a/f_2$),
trajectories in the \pom -proton interaction,
the \pom -\pom -Reggeon process (${\cal P}{\cal P}{\cal R}$).
The latter term is largest at low energies and decreases with increasing $s$.
We shall see below that there is good evidence that the 
${\cal P}{\cal P}{\cal R}$ term is required by the data.
Donnachie and Landshoff\cite{dl_tot} determined the ``effective''
$\epsilon$ values to be 0.08 and -0.45, respectively,  
from fits to the $s$-dependence of available total cross section data.
However, recent analyses\cite{cudell,dino2} and new data yield somewhat 
different values and, in the fits reported 
here, 
we use\footnote{We do not have sufficient constraints with our data
to allow $\epsilon_1$ and $\epsilon_2$ to be free parameters.} 
$\epsilon_1 = 0.10$ and $\epsilon_2 = -0.32$.
$R$ is a free parameter.
\item The product, $K \cdot \sigma_0$, is the magnitude of
$G_{{\cal P}{\cal P}{\cal P}}$, and the product, $K \sigma_0 R$, 
is the magnitude of $G_{{\cal P}{\cal P}{\cal R}}$. 
Since the overall normalization constant, $K$, of \flux\
is not uniquely defined theoretically,
and since it multiplies \sigpomtot\ in Eq.~\ref{eq:tripleR}, 
only the product, $K \cdot \sigma_0$ is measurable. 
This is one of the free parameters in our fits.
\item
The ``standard" \pom\ Regge trajectory, determined from low momentum transfer
data, is linear: $\alpha (t)= 1 + \epsilon + \alpha ' t $, with the 
slope\cite{dl_elastic}: $\alpha ' = 0.25$~GeV$^{-2}$.
However, we show in the next section that this linear trajectory
is too small in the 1--2~GeV$^2$ $|t|$ -region of this experiment to adequately 
describe the data, and we allow for a quadratic term in the trajectory:
\begin{equation}
\alpha (t) \, \, = \, \, 1.10 \, \, + \, \, 0.25 t \, \, + \, \, \alpha '' t^2
\label{eq:alpha}
\end{equation}
\item The multiplicative factor , $e^{bt}$, is found to be required by the data
to compensate for the presence of 
the quadratic component in the \pom\ trajectory.
The introduction of this factor is not sacrilegious, because $|F_1(t)|^2$ has 
never been shown to describe React.~\ref{eq:dif} at large $|t|$. 
Although
Donnachie and Landshoff have suggested\cite{dl_dif} that \sigpomtot\ may
also depend on momentum transfer, \T , we ignore that possibility in
this paper but note that any such dependence would be absorbed in the
$e^{bt}$ factor.

\end{itemize}

\subsection{Measurement of the effective {\boldmath \pom\ } trajectory}
\label{peakfits} 
\indent
 
In order to determine the effective \pom\ trajectory at given values of
\T , we fit to the uncorrected data points with 
$\xp > 0.97$ in Figs.~\ref{fig:dndx}.
The appropriate function is found by noting that, at a fixed momentum transfer 
\T , Eq.~\ref{eq:tripleR} can be written as: 
\begin{equation}
{{d^2 \sigma}\over{d \xi d t}}\, \, \propto \, \, \xi^{1 - 2 \alpha (t)} 
\x \sigpomtot (\xi s)
\label{eq:xim1}
\end{equation}
where we recall that $s' = \xi s$. 
It is evident that the explicit functional dependence of 
\sigpomtot\ on $\xi$ plays a significant role in the fits.
With the use of Eq.~\ref{eq:twocomp}, we rewrite Eq.~\ref{eq:xim1} as:
\begin{equation}
{{d^2 \sigma}\over{d \xi d t}}\, \, \propto \, \, 
\xi ^{1-2 \alpha (t) + \epsilon_1} \, \, + 
\, \, R \x s^{\epsilon_2 - \epsilon_1} \x \xi ^{1-2 \alpha (t) + \epsilon_2}
\label{eq:xim2}
\end{equation}
\begin{equation}
{{d^2 \sigma}\over{d \xi d t}}\, \, \propto \, \, 
\xi ^{1.10 - 2 \alpha (t)} \, \, + \, \, 
0.0178 \, \x \, \xi ^{0.68 - 2 \alpha (t)} 
\label{eq:xim3}
\end{equation}
In evaluating the constants in Eq.~\ref{eq:xim2}, we assume $R$ = 4.0 (see
Sect.~\ref{combined}), $\epsilon_1 = 0.10$ and 
$\epsilon_2 = -0.32$, leading to Eq.~\ref{eq:xim3}.
Fitting this function to data in the $\xp \sim 1$ peak region in a given
\T -bin yields the value of $\alpha$ at that \T -value.
To perform the fits, Monte-Carlo events
were generated according to the function, then multiplied by the acceptance 
function and offline selection efficiencies
and smeared according to the experimental resolution.

The traditional fit to the differential cross section in the peak
region, using the function $\xi ^{m}$, ignores the 
${\cal P}{\cal P }{\cal R }$ term in Eq.~\ref{eq:xim2},
which is negligible at large $\xi$ and $s$.
However, in fits at small $\xi$, where the second term is significant,
we have performed our fits both with and without it.
We note that the fits themselves do not distinguish between the
two cases. The evidence that the data require both terms is given in 
Sect.~\ref{combined}. 

\begin{figure}
\begin{center}
\mbox{\epsfig{file=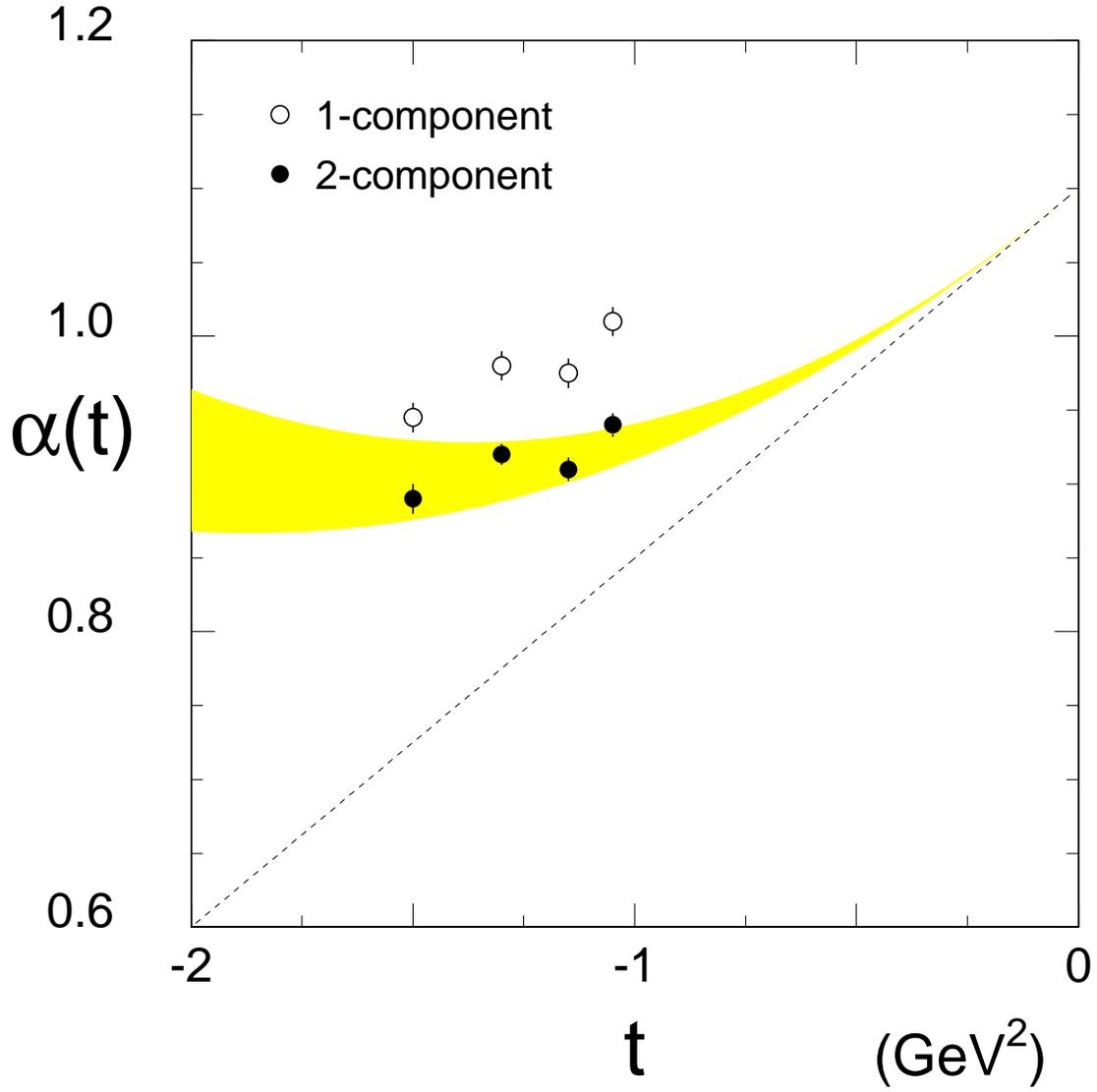,width=16.0cm}}
\end{center}
\caption[]{
The \pom\ $\alpha (t)$ points are from fits to the small-$\xi$ peak regions 
shown in Fig.~\protect\ref{fig:dndx_fit}, as described in the text. 
The dashed curve is the linear trajectory: $\alpha (t) = 1.10 + 0.25 t$.
The shaded band shows the effect of adding a quadratic term, 
$\alpha '' t^2 = 0.079 t^2$, to the \pom\ trajectory in the fits
to the data in the $\xi$-range, 0.03-0.10, described in the text.
The width of the shaded band shows the 
$\pm 1 \sigma$ error range on $\alpha ''$.
}
\label{fig:alpha}
\end{figure}

\begin{figure}
\begin{center}
\mbox{\epsfig{file=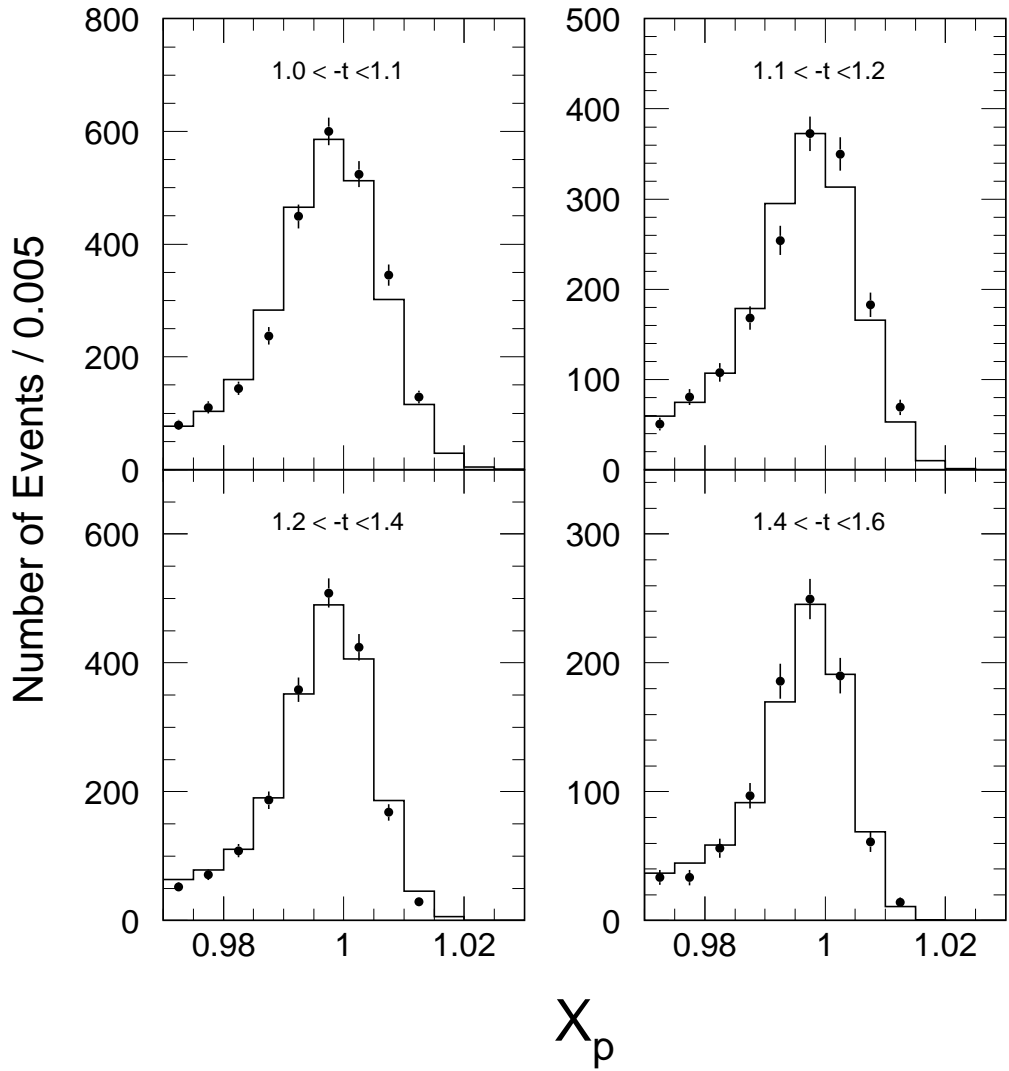,width=16.0cm}}
\end{center}
\caption[]{
Fits (histograms) 
of Eq.~\ref{eq:xim2}, using both components, to the uncorrected
distributions (points) seen in Fig.~\ref{fig:dndx}.
The inclusion of acceptance and resolution in the fits is
described in the text. $t$ is given in units of GeV$^2$.
}
\label{fig:dndx_fit}
\end{figure}

Table~\ref{tab:mvalues} contains values of $\alpha (t)$ obtained 
with and without the ${\cal P}{\cal P }{\cal R }$ term. 
The resulting $\alpha (t)$ values for both types of fits
are plotted vs. \T\ in Fig.~\ref{fig:alpha}. 
The fits using the 2-component \sigpomtot\ 
are shown in Fig.~\ref{fig:dndx_fit}.
We have fit only in the range, $1.0 < -t < 1.6$ GeV$^2$, where our
experimental resolution is best understood.

\begin{figure}
\begin{center}
\mbox{\epsfig{file=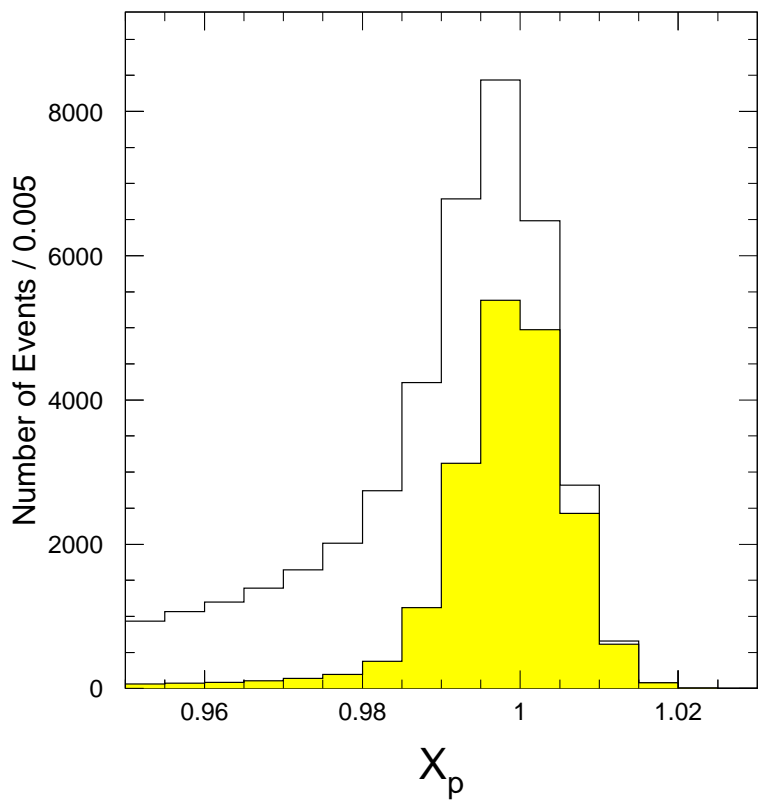,width=12.0cm}}
\end{center}
\caption[]{
The two terms of Eq.~\ref{eq:xim2} fitted to the data set in 
Fig.~\ref{fig:dndx_fit} with $1.2 < -t < 1.4$ GeV$^2$.
The ${\cal P }{\cal P}{\cal P}$ term is open.
The shaded distribution superimposed on it is the
${\cal P}{\cal P }{\cal R }$ term.
As discussed in the text, the fits assume $R$ = 4.0.

}
\label{fig:dndx_mc}   
\end{figure}

The $\alpha$ values obtained with the 2-component cross section are 
seen to be systematically lower by about 0.05 than those obtained with the
1-component cross section. This can be understood with reference to
Fig.~\ref{fig:dndx_mc}, which shows the two components in one of 
the fits plotted separately. 
Since the second component is largest at small $\xi$, its
\xp\ distribution drops off much more rapidly on the low side of the peak.
Thus, a fit without the second component included requires a larger 
value of $\alpha$ to fit to the same shape data distribution.

The shaded region
in Fig.~\ref{fig:alpha} results from the global fits to all UA8
and ISR data in the $\xi$-range, 0.03--0.09, as described in
Sec.~\ref{combined}.
We note that the (2-component) 
points and the shaded band are in essential agreement, even
though they are obtained by analyzing different data with different methods.
They demonstrate that the \pom\ trajectory departs from linearity in our
\T -region and is substantially larger than the linear function.
Frankfurt and Strikman\cite{fs} and 
Collins et al.\cite{collins} have pointed out
that such an increasing effective trajectory may 
arise from the onset of the perturbative two-gluon \pom . 
Another possibility\cite{kaidalov,capella} 
is that the exchange of two \pom s (or more) might be 
responsible for such an effect. It is well known that two-\pom -exchange
plays an important role in understanding elastic scattering\cite{dl_elastic}.

\subsection{$s$-Dependence of {\boldmath \dsig\ at fixed $\xi$ and $t$}}
\label{sec:sdep}
\indent

From the differential cross section, Eq.~\ref{eq:tripleR}, we see that,
at fixed $\xi$ and \T , and because $s' = \xi s$, 
the differential cross section directly displays
the $s$-dependence of $\sigpomtot (s')$    
multiplied by any possible $s$-dependence of $K$\cite{dino}.  
\begin{equation}
{{d^2\sigma}\over{d\xi dt}}\, \, \propto \, \, K \, \x \, \sigpomtot (\xi s)
\label{eq:fixedxit}
\end{equation}

We first consider the $\xi$-range, 0.03--0.04:
\begin{itemize}
\item $\xi > 0.03$ is sufficently far from 
the large peak near $\xi \sim 0$, that resolution smearing does not
distort the observed distribution.
\item Background from non-\pom -exchange is small, 15\% or less 
in this region\cite{sens} and is the size of the
data points on the plots. Thus, for our initial fits, we can ignore it. 
The analysis of the full data set, 
as described in Sect.~\ref{combined} is consistent with a background of
this magnitude.
\end{itemize}

\begin{figure}
\begin{center}
\epsfig{file=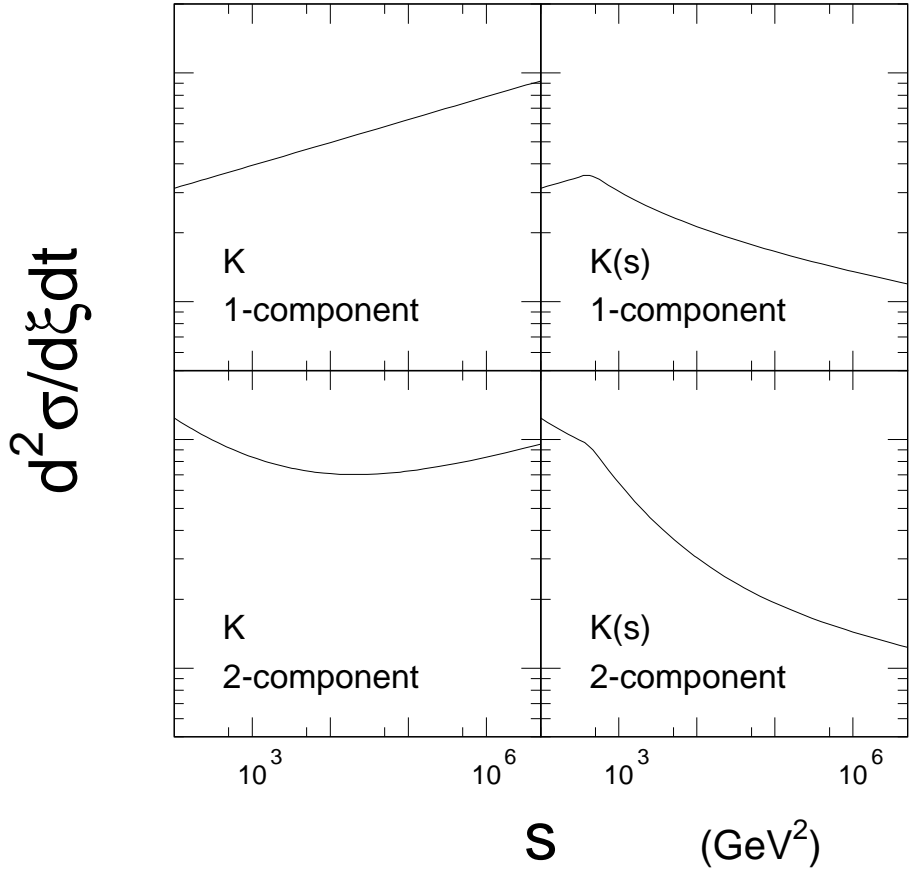,width=13cm}
\caption[]{
Expected shapes of $s$-dependences of \dsig\ at fixed \T\ and $\xi$
for the four different combinations
of possibilities, with and without an $s$-dependent $K$, and with and
without the second term in \sigpomtot . Although the curves are calculated 
at $-t$ = 1~GeV$^2$ using
the results of Fit ``A" (see Table~\ref{tab:chisqbknd}), 
their essential properties are the same at all \T\ values.
}
\label{fig:dsdxdt_shape}
\end{center}
\end{figure}

Figure~\ref{fig:dsdxdt_shape} shows what is expected for the $s$-dependence
of \dsig\ for the four combinations of possibilities, with and without
an $s$-dependence of $K$, and with and without the 
${\cal P}{\cal P }{\cal R }$ term in \sigpomtot . The behaviors seen in
Fig.~\ref{fig:dsdxdt_shape} are easily understood:
\begin{itemize}
\item 1-component \sigpomtot\ and constant $K$: \, \, \dsig\ increases uniformly
according to $s^{0.10}$
\item 2-component \sigpomtot\ and constant $K$: \, \, \dsig\ initially decreases
because of the $s^{-0.32}$ term, and then increases because of the 
$s^{0.10}$ term.
\item 1-component \sigpomtot\  and $K(s)$: \, \, \dsig\ initially increases
according to $s^{0.10}$, and then decreases above 
$\sqrt{s} = 22$~GeV because of the ``renormalization" of $K$\cite{dino}.
\item 2-component \sigpomtot\ and $K(s)$: \, \, \dsig\ initially decreases
because of the $s^{-0.32}$ term. Above $\sqrt{s} = 22$~GeV,
the presence of the decreasing term coupled with 
the ``renormalization" of $K$ leads to a rapidly decreasing behavior.
\end{itemize}
It is evident that data should easily distinguish between the four cases.

\begin{figure}
\begin{center}
\epsfig{file=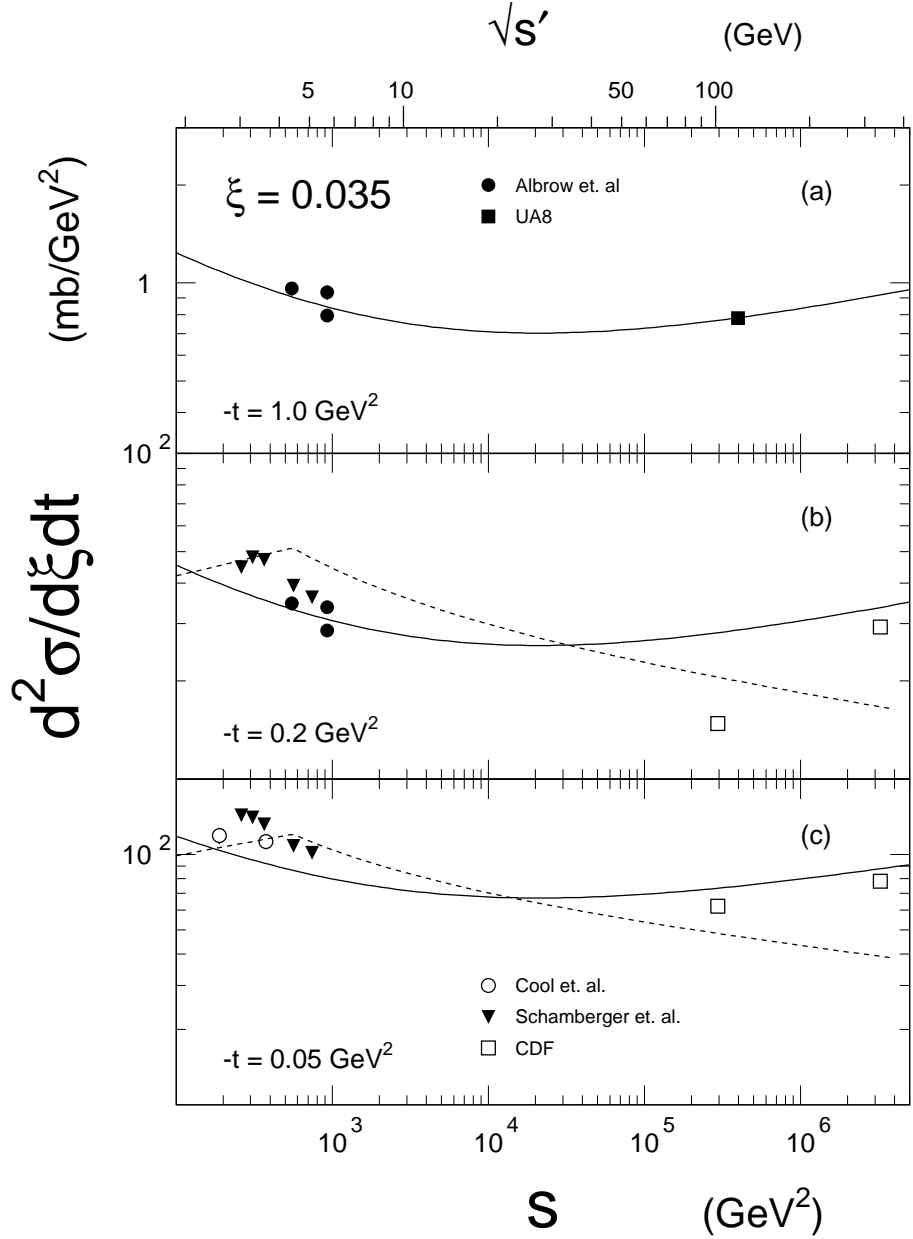,width=13cm}
\caption[]{
\dsig\ vs. $s$ at fixed $\xi = 0.035$ and at three different \T -values. 
The solid curves are calculated from Fit ``A" in Table~\ref{tab:chisqbknd}.
The dashed curves are calculated using the renormalized flux factor
of Ref.~\protect\cite{dino}.
The references are: 
Albrow et al.\protect\cite{albrow},
Schamberger et al.\protect\cite{schamberger},
Cool et al.\protect\cite{cool}
and the CDF Collaboration\protect\cite{cdf}.
See the discussion of the CDF points in the text.
}
\label{fig:sdep}
\end{center}
\end{figure}

Figures~\ref{fig:sdep}(a-c) show the available
experimental values of \dsig\ vs. $s$ 
at $\xi = 0.035$ and at three fixed values of \T .
The weak dependence on $s$ of the $-t$ = 1~GeV$^2$ ISR and UA8 data points 
in Fig.~\ref{fig:sdep}(a)
reflects what we have already seen in Fig.~\ref{fig:compisr}.
The solid curves are triple--Regge predictions for case (ii)
and are discussed in the following section.

\begin{figure}
\begin{center}
\epsfig{file=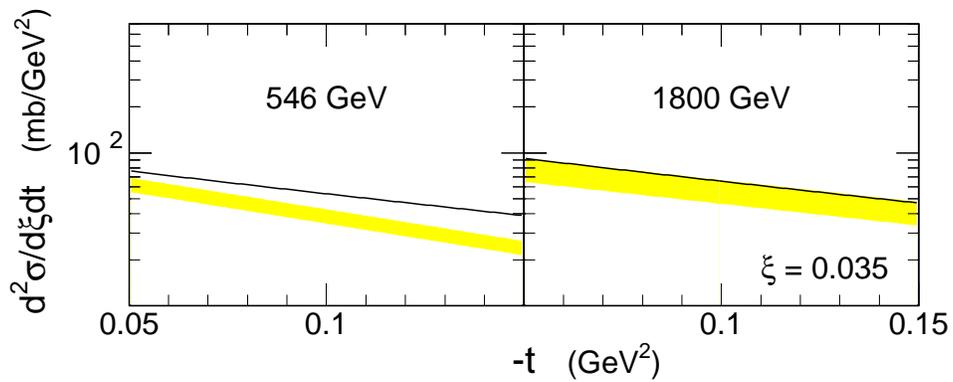,width=15cm}
\caption[]{
Bands are the CDF differential cross sections calculated at $\xi = 0.035$
from their fitted functions; widths are $\pm$1$\sigma$ error on the
amplitudes (as explained in the text, ``signal" and ``background" are
added together). The curves are from the same calculations used
for the solid curves in Fig.~\ref{fig:sdep}.
}
\label{fig:cdft}
\end{center}
\end{figure}

At the two lower $|t|$ values, which
approximately span the region where CDF has their data,
points from the CDF collaboration\cite{cdf} are 
calculated from their fitted 
function\footnote{However, since their quoted background levels
are larger than in other experiments, 
where the background in the $\xi$-region shown is small enough to be ignored,
we show the sum of
their fitted ``signal" and ``background".}.
See also Fig.~\ref{fig:cdft}.
We note that the two CDF points
are not self-consistent (their fitted \T -slopes are ($7.7 \pm 0.6$) GeV$^{-2}$
and ($4.2 \pm 0.5$) GeV$^{-2}$, at $\sqrt{s} = 546$~GeV and 1800 GeV, 
respectively; the latter is in 
excellent agreement with $|F_1 (t)|^2$ in Eq.~\ref{eq:tripleR}.
However, we note that,
at $|t|$ = 0.05~GeV$^2$, where the different $|t|$-slopes have much
less effect, the 546 and 1800~GeV data display
the increase with energy predicted by the solid curve.

We also note that, at low $|t|$ in Figs.~\ref{fig:sdep}(b,c),
the Schamberger et al. points\cite{schamberger}
are systematically larger by 10--15\% than {\it both} the Cool et 
al.\cite{cool} and Albrow et al.\cite{albrow} points.
Nonetheless, it is evident from a comparison of 
Fig.~\ref{fig:dsdxdt_shape} with Fig.~\ref{fig:sdep} and \ref{fig:cdft}, 
that the data are inconsistent
with all cases where \sigpomtot\ has  a single component,
and/or where
$K(s)$ follows the ``renormalization" procedure of Goulianos\cite{dino}.

\subsection{Combined fits to UA8 and ISR data}
\label{combined}
\indent

\begin{figure}
\begin{center}
\epsfig{file=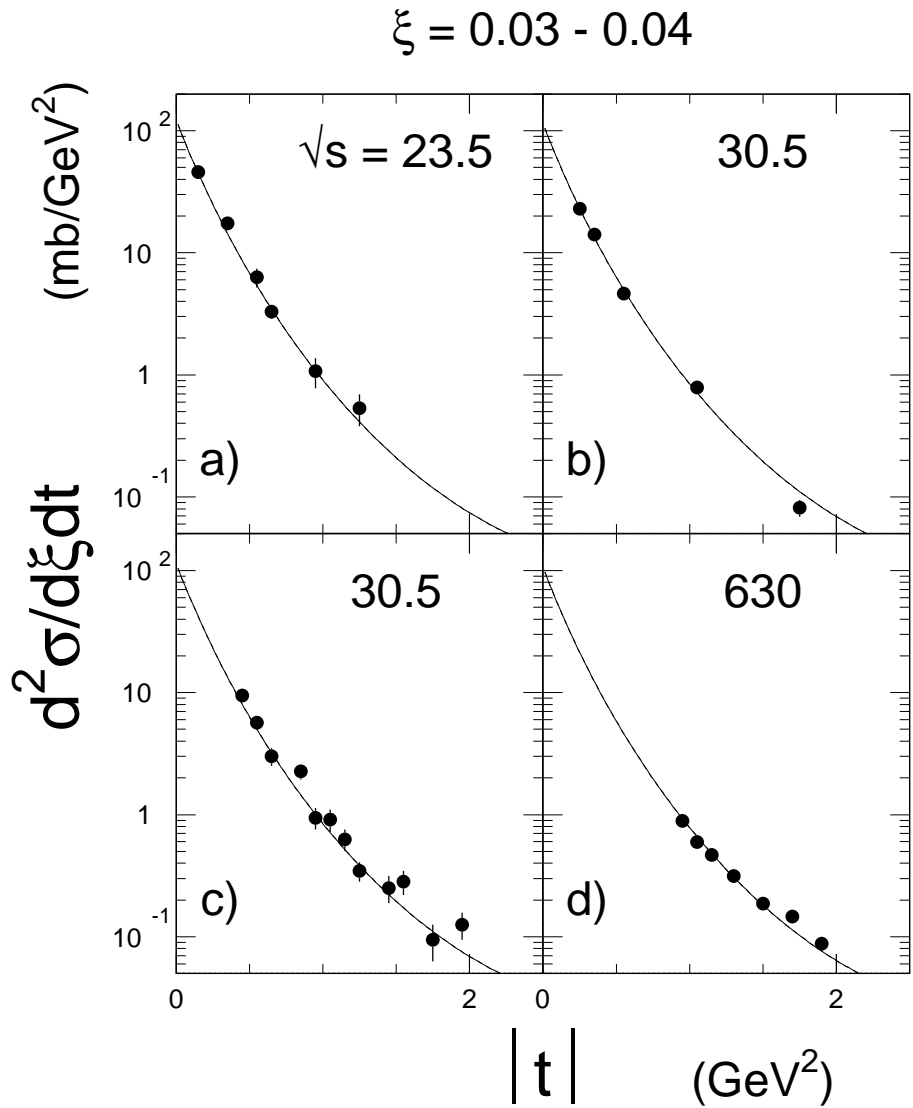,width=13cm}
\caption[]{
Differential cross section, \dsig , vs. $|t|$ , for 3 ISR 
measurements\protect\cite{albrow}
and UA8. Where more than one data point exists in the $\xi$ interval,
0.03--0.04, their average is shown on this plot; thus, 30 points are shown,
although 48 were used in performing Fit ``A".
The curves correspond to Fit ``A" in Table~\ref{tab:chisqbknd} evaluated
at $\xi = 0.035$. 
}
\label{fig:dsdt035}
\end{center}
\end{figure}

To test the validity of Eqs.~\ref{eq:tripleR} and~\ref{eq:twocomp}
we fit our UA8 data simultaneously with 
the ISR data\cite{albrow}.
In our first fit, we continue to focus on 
the narrow $\xi$-range, 0.03--0.04, where experimental 
resolution and non-\pom -exchange background issues 
do not play a significant role and are ignored in our analysis.
Figure~\ref{fig:dsdt035} shows the mean \dsig\ in the range $\xi$ = 0.03--0.04 
plotted vs. $|t|$, for each of four data sets (there are two sets of
ISR measurements at $s = 930$~GeV$^2$).

Table~\ref{tab:chisq} shows the results of fitting Eqs.~\ref{eq:tripleR} and 
\ref{eq:twocomp} to all available data points 
in the $\xi$ = 0.03--0.4 range,
with various combinations of free parameters in the fits. 
It is clear that, in order to have an acceptable $\chi ^2$/Degree of Freedom,
$\alpha ''$, $b$ and $R$ are all required. 
The last fit in the table, in which all
three parameters are included, is shown superimposed on the data in
Figs.~\ref{fig:dsdt035} and is seen to be quite good.

One of the remarkable results of the fit is that $R \neq 0$. 
This shows that the
${\cal P}{\cal P }{\cal R }$ term is required by the data, as discussed
in the previous section. 

A second remarkable result of the fit is that the value of
$\alpha ''$ = ($0.078 \pm 0.013$)~GeV$^{-4}$ 
leads to a \pom\ trajectory, plotted as the shaded band in
Fig.~\ref{fig:alpha}, which is in good agreement with the values 
obtained from the 2-component fits to the 
small-$\xi$ peaks in Figs.~\ref{fig:dndx_fit}.
This result reinforces our conclusion that the \pom\ trajectory
flattens in our \T -range and also lends credence to the other aspects of the
fit to the $\xi$ = 0.03--0.04 data. 
Finally, we note that although there is a strong correlation between the values
of $\alpha ''$ and $b$, the best fit is obtained when both are non-zero.

The fit leads to definite predictions
for the s-dependence of \dsig\ at fixed $\xi$ and \T , 
as discussed in the previous section.
The solid curves in Figs.~\ref{fig:sdep} and \ref{fig:cdft}
are calculated using the quoted
convergence values of the free parameters and,
not unexpectedly, in Fig.~\ref{fig:sdep}
passes through the UA8 point at $-t$ = 1.0 GeV$^2$
and the averages of the ISR points at $-t$ = 1.0 and 0.20 GeV$^2$
which were used in the fit. 
However, because the CDF points points are 
are outside our fitted range of $s$ and, in particular,
$-t$ = 0.05~GeV$^2$ is below our fitted range of \T ,
it is remarkable that they are in reasonable agreement 
with the solid curves.

The dashed curves are Eq.~\ref{eq:tripleR}, evaluated following
the prescription of Goulianos\cite{dino} (see Sect.~\ref{totsig}).
As discussed above, they do not agree with the data.
If the renormalization hypothesis alone is discarded, then the prediction 
follows the linear segment of the dashed curve over the entire range of $s$ 
shown in the figure, in disagreement with the data. 
It is thus clear that the one-component \sigpomtot\ hypothesis must also be 
incorrect. 
If the renormalization hypothesis is retained, then \sigpomtot\ would have to 
increase much more rapidly with $s$ than it does. 

We now turn to the larger $\xi$-region, 0.03--0.09, where non-\pom -exchange
background must be included.
This background is described by adding an empirical background 
term\cite{sens} 
to Eq.~\ref{eq:tripleR} of the form: 
\begin{equation}
({{d^2\sigma}\over{d\xi dt}})_{Background} \, \, = \, \, A \xi ^1 e^{ct}, 
\label{eq:back}
\end{equation}
with different values of $A$ and $c$ for the ISR data and the UA8 data.
In all fits described below where data in the larger $\xi$-range, 0.03--0.09,
are used, the background term is applicable.

We have also attempted to describe the background by adding 
other Reggeon terms in the triple-Regge formula,
but could not achieve acceptable fits. This is
perhaps not surprising, since we know that there are other types of 
background. For example, it is known that 10\% of all protons at $\xi = 0.1$
come from large-\xp\ production of $\Delta ^{++}$\cite{bc300}. 
There can also be other types of experiment-dependent backgrounds.

Table~\ref{tab:chisqbknd} shows the results of all simultaneous fits 
made to the UA8 and ISR data\cite{albrow} 
in the larger $\xi$-range, 0.03--0.09.
Fit ``A" is the fit to the $\xi$ = 0.03--0.04 data, 
described above. Fit ``B" adds to these data all the
UA8 data up to $\xi = 0.09$. Fit ``C" does the same as Fit ``B" except
that the ISR data and not the UA8 data are added. Fit ``D" adds both to the
$\xi$ = 0.03--0.04 data.  

Although, when comparing Fits ``B" and ``C", we see some evidence of systematic
shifts of the parameters, the errors do tend to overlap in the two 
cases.
Of more significance is the fact that the results from 
Fit ``D'' are seen to agree rather well with Fit ``A". In particular,
$K \sigma _0$,  $\alpha ''$, b and R in the two cases are consistent within
the quoted statistical uncertainties.
The fact that the two fits yield self-consistent results
is a good indication that they are reliable. 
We believe that Fit ``D" is the best available description of the experimental
differential cross section, \dsig .

It is particularly interesting to note that the value $R$ = 4.0 $\pm$ 0.6, 
is comparable to the $\sim$3.5 value found in the 
fits to real $pp/p\bar{p}$ total cross sections\cite{cudell,dino2}. 
This seems to say that the relative strengths of \pom\ and $a/f_2$-exchange
are nearly the same in proton-proton and \pom -proton total cross 
sections.

\begin{figure}
\begin{center}
\mbox{\epsfig{file=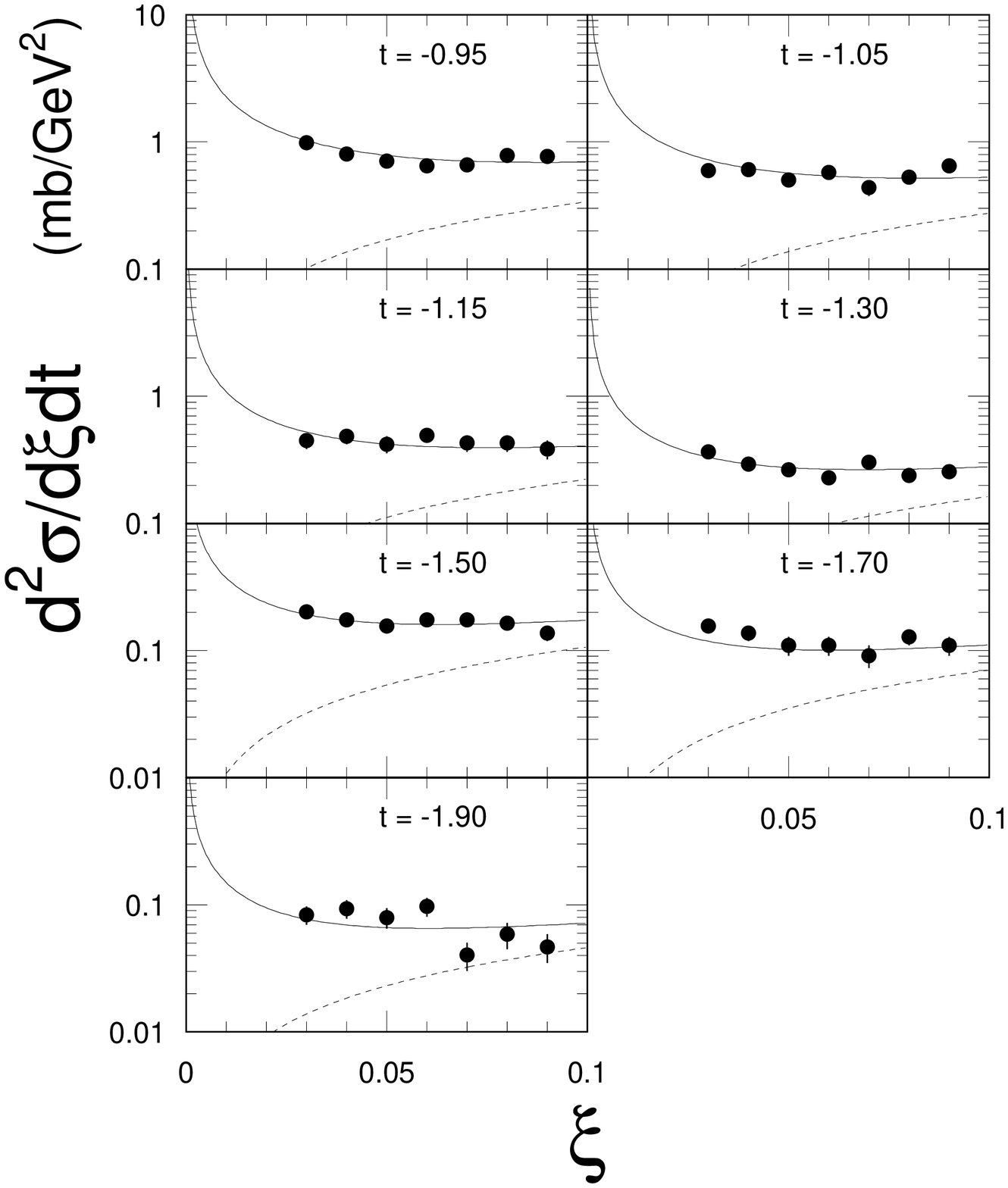,width=13cm}}
\end{center}
\caption[]{
\dsig\ vs. $\xi$ for the indicated seven bins of \T\ (given in units of 
GeV$^2$). 
In each case, the solid curve is the fitted function 
given by the sum of Eqs.~\ref{eq:tripleR} and \ref{eq:back} using Fit ``D".
The same fits are shown in Figs.~\ref{fig:samimtt}. 
The solid curves include the non-\pom -exchange background from the fits.
The dashed curves are the background alone.
}
\label{fig:samimxi}
\end{figure}

\begin{figure}
\begin{center}
\mbox{\epsfig{file=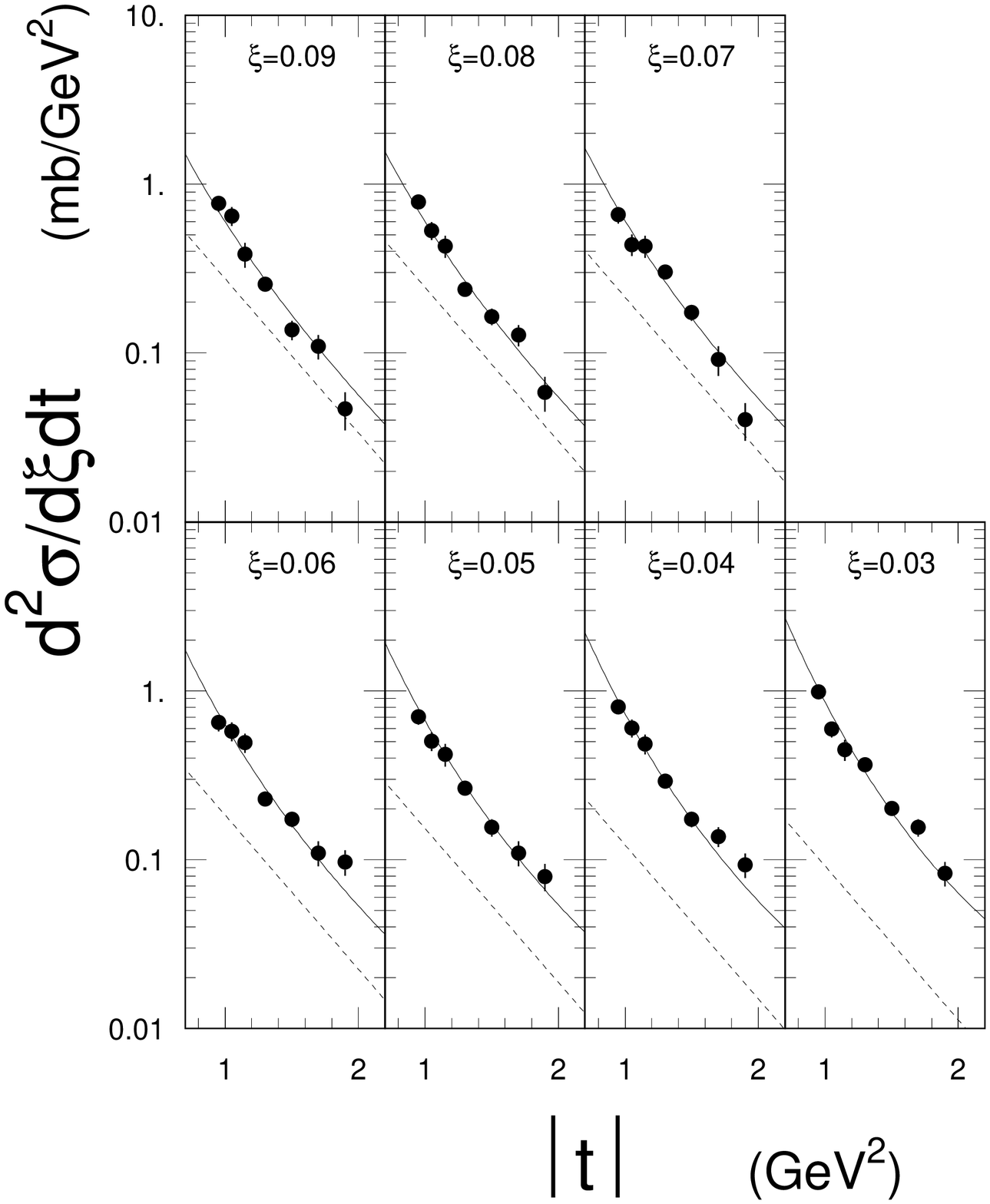,width=13cm}}
\end{center}
\caption[]{
\dsig\ vs. \T\ for the indicated seven bins of $\xi$. 
In each case, the solid curve is the fitted function given by the sum of 
Eqs.~\ref{eq:tripleR} and \ref{eq:back} using Fit ``D".
The same fits are shown in Figs.~\ref{fig:samimxi}. 
The solid curves include the non-\pom -exchange background from the fits.
The dashed curves are the background alone.
}
\label{fig:samimtt}
\end{figure}

\begin{figure}
\begin{center}
\epsfig{file=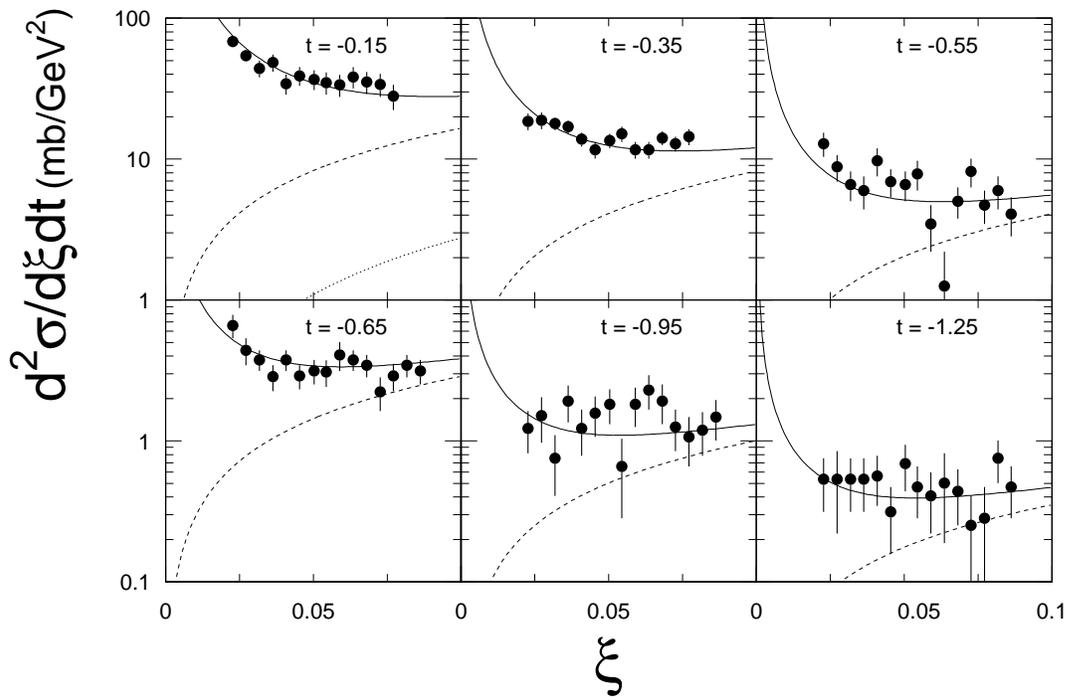,width=15cm}
\caption[]{
ISR data with $s = 551$~GeV$^2$\protect\cite{albrow}. 
In each case, the solid curve is the fitted function given by the sum of 
Eqs.~\ref{eq:tripleR} and \ref{eq:back} using Fit ``D".
Only points with $\xi > 0.03$ are used in the fit.
The solid curves include the non-\pom -exchange background from the fits.
The dashed curves are the background alone.
The dotted curve on the $-t = 0.15$~GeV$^2$ plot is the one-pion-exchange
contribution to the non-\pom -exchange background described in the text
(not included in the fit).
}
\label{fig:isr551xx}
\end{center}
\end{figure}

\begin{figure}
\begin{center}
\epsfig{file=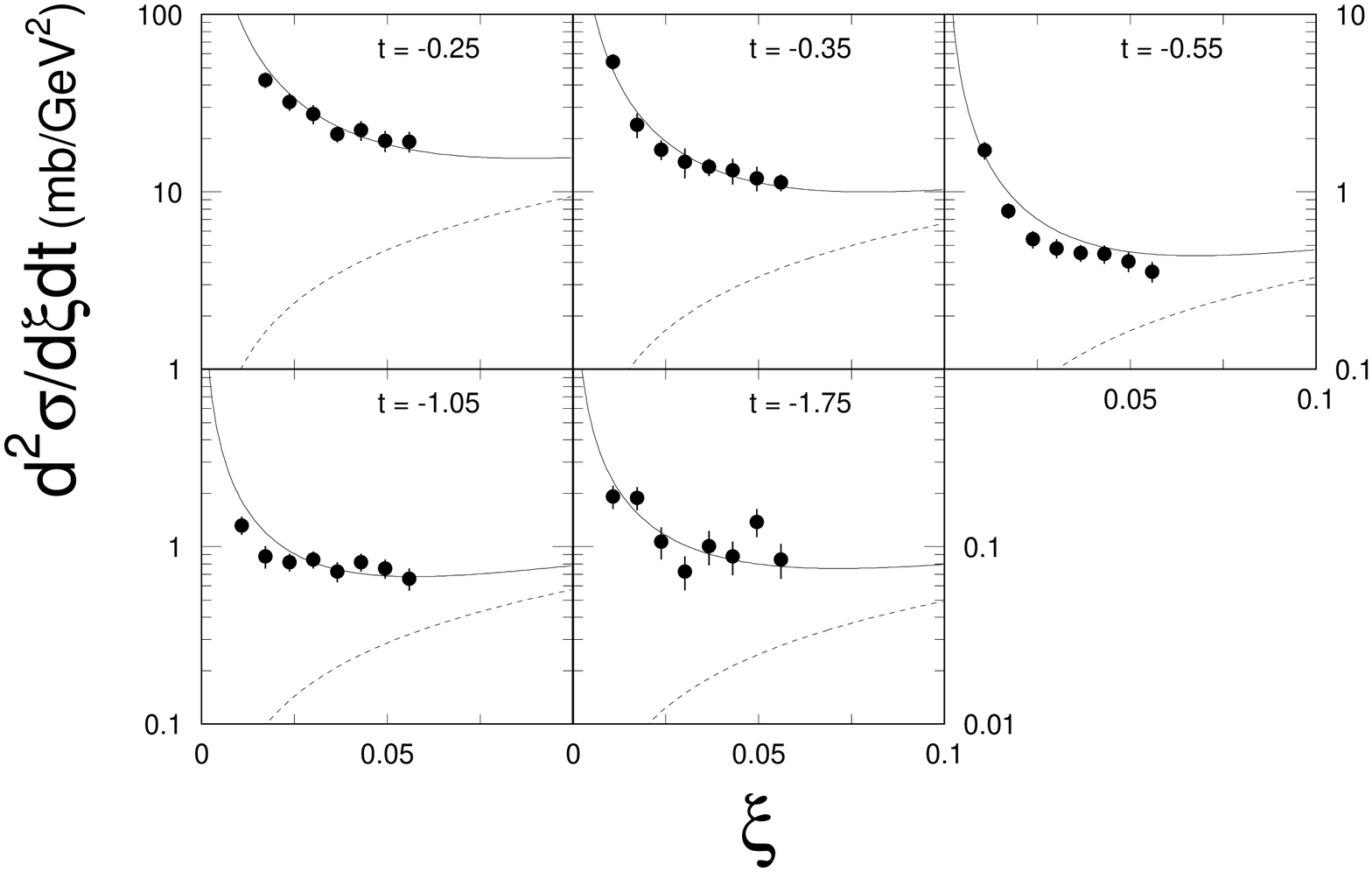,width=15cm}
\caption[]{
ISR data with $s = 930$~GeV$^2$\protect\cite{albrow}. 
In each case, the solid curve is the fitted function given by the sum of 
Eqs.~\ref{eq:tripleR} and \ref{eq:back} using Fit ``D".
Only points with $\xi > 0.03$ are used in the fit.
The solid curves include the non-\pom -exchange background from the fits.
The dashed curves are the background alone.
}
\label{fig:isr930xx}
\end{center}
\end{figure}

\begin{figure}
\begin{center}
\epsfig{file=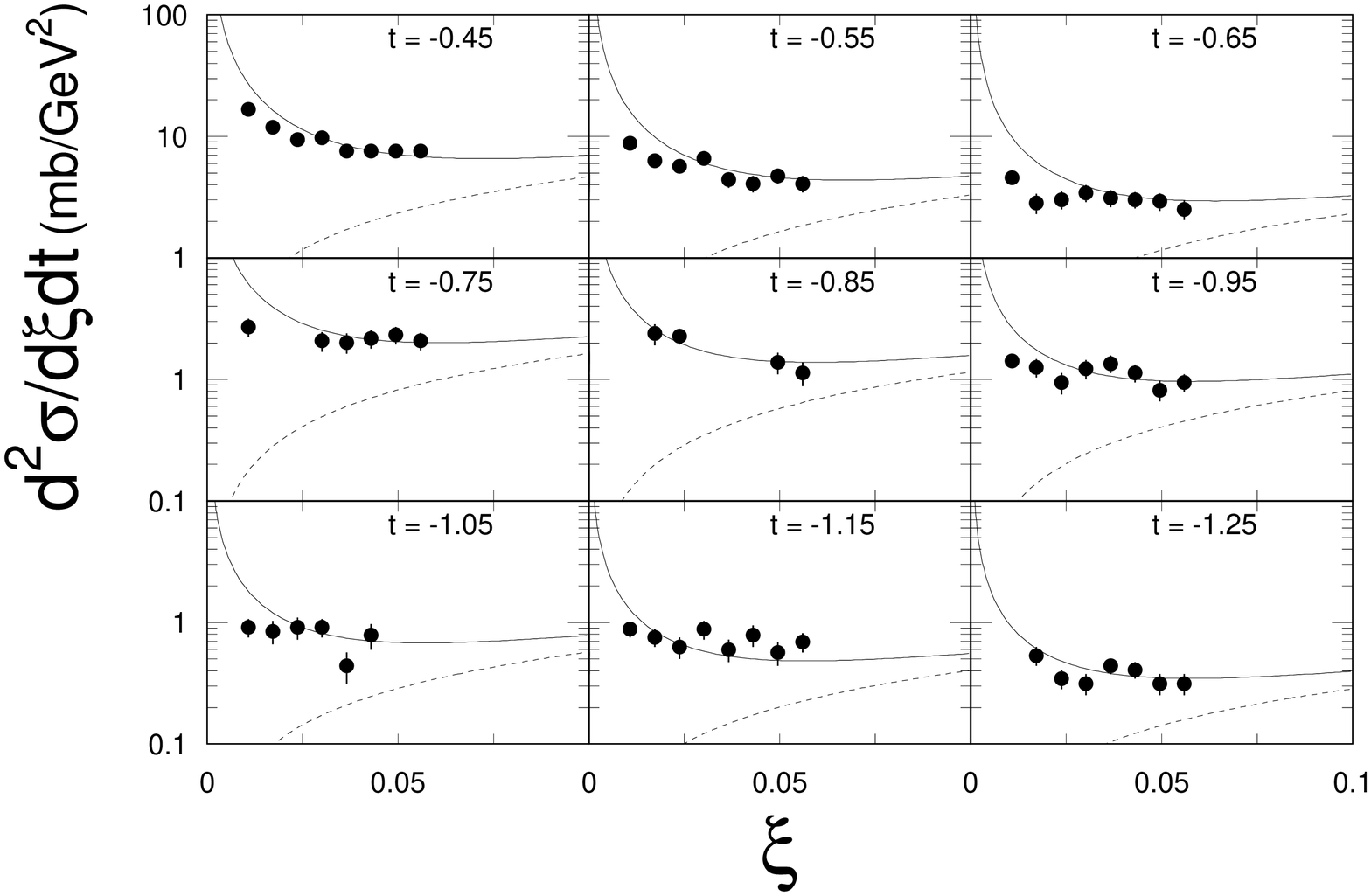,width=15cm}
\caption[]{
A second set of ISR data with $s = 930$~GeV$^2$\protect\cite{albrow}.
In each case, the solid curve is the fitted function given by the sum of 
Eqs.~\ref{eq:tripleR} and \ref{eq:back} using Fit ``D".
Only points with $\xi > 0.03$ are used in the fit.
The solid curves include the non-\pom -exchange background from the fits.
The dashed curves are the background alone.
}
\label{fig:isr931xx}
\end{center}
\end{figure}

Figures~\ref{fig:samimxi}, \ref{fig:samimtt}, 
\ref{fig:isr551xx}, \ref{fig:isr930xx} and \ref{fig:isr931xx}
display most of the data used in the fits described in this section.
Superimposed on the data points are the fitted curves (using Fit ``D")
for signal plus background and background alone. The curves are seen to
be in reasonably good agreement with the data used.

\subsection{One-pion-exchange}
\label{ope}
\indent

Since one-pion-exchange (OPE) is a well-established phenomenon, we consider
to what extent it can account for the non-\pom -exchange background found 
in our fits. The exchanged particle in the diagram of 
Fig.~\ref{fig:difdiag} would then be a $\pi ^0$.

At high energies, large $s'$ and low $|t|$, 
the Chew-Low equation\cite{chewlow} can be 
written in terms of a ``$\pi ^0$ flux factor in the proton\cite{schlein},
$F_{\pi^0/p} (t, \xi)$ (i.e., for a single-arm in $pp$-collisions):
\begin{equation}
\label{eq:ope}
{{d^2 \sigma }\over{d \xi dt}} \, \, = \, \,  F_{\pi^0/p}(t, \xi) \, 
\x \, \sigma_{\pi^0 p}(s')
\end{equation}
\[ F_{\pi^0/p}(t, \xi) = {{14.6}\over{4 \pi}} \, \,
{|t|\over{(t- \mu^2)^2}} \, \, \xi ^1 \, \, [F_{ope}(t)]^2 \]
where, as in Eq.~\ref{eq:tripleR} for the \pom , the squared mass at the upper
vertex, $s' = \xi s$ 
($\xi$ is now the pion's momentum fraction in the proton).
The factor, 14.6, is the $\pi ^0$-nucleon coupling constant at the
lower vertex in Fig.~\ref{fig:difdiag} with an exchanged pion, 
$\mu$ is the pion mass and
$\sigma_{\pi ^0 p}(s')$ is the $\pi^0$-proton total cross section 
($\sim 29$~mb). 
In contrast with the $\sim \xi^{-1}$ dependence for the \pom ,
the $\xi^1$ dependence for the pion reflects the fact that
$\alpha_\pi (t) \sim 0$ at low $|t|$ (and that
the pion contains a beam-proton valence quark). 

$F_{ope}(t)$ is an empirical form factor which was shown in the early days
of OPE studies, 30 years ago, to extrapolate correctly to on-shell 
cross sections at the pion-exchange pole in the scattering amplitude.
Its form is given in Ref.~\cite{ma}.

Using Eq.~\ref{eq:ope} and the form factor from Ref.~\cite{ma}, the 
dotted curve at $-t = 0.15$ in Fig.~\ref{fig:isr551xx} shows
the expected OPE contribution to the observed cross section. 
We see that the OPE background is only a small part of the total 
non-\pom -exchange background.

\subsection{Breakdown of the triple-Regge formula in the resonance region}
\label{breakdown}
\indent

It is interesting to investigate how low in $s'$ and $s$ the triple-Regge
formula describes data. To determine this, we note that,
in the absence of non-\pom -exchange background,
Eq.~\ref{eq:xim1} at fixed \T\ can be rewritten as:
\begin{equation}
\label{eq:snorm}
(s)^{1-2\alpha (t)} \, \, \x \, \, {{d^2 \sigma }\over{d \xi dt}} 
\, \, \, \propto \, \, \, (s')^{1-2\alpha (t)} \, \, \x \, \, \sigpomtot (s')
\end{equation}
Thus, a plot of the differential cross section multiplied by
$s^{1-2\alpha (t)}$ vs. $s'$ should be independent of $s$.

\begin{figure}
\begin{center}
\epsfig{file=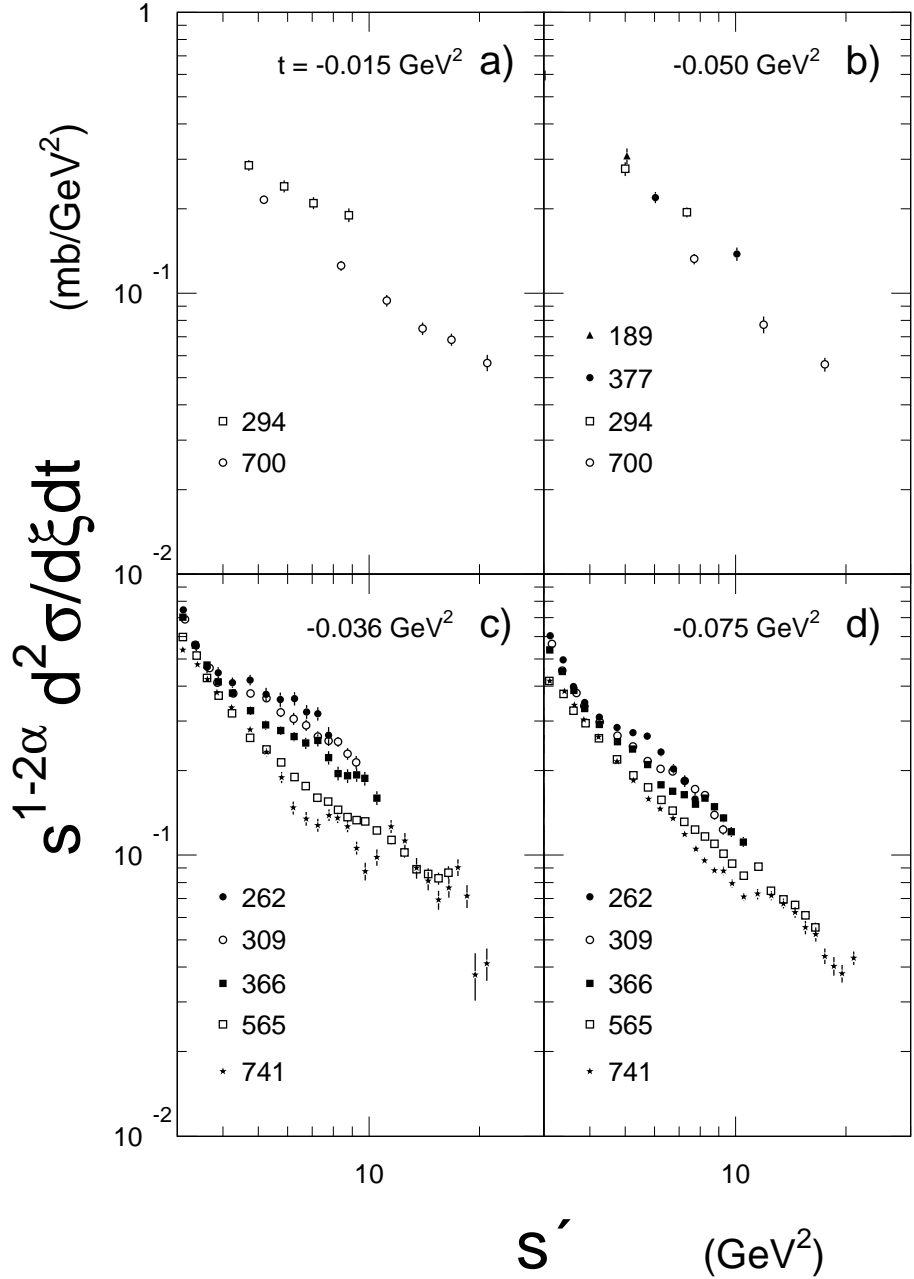,width=13cm}
\caption[]{
The quantity, $s^{1-2\alpha (t)} \x \dsig$ vs. $s'$ at four
values of \T . 
The $s$ values (in GeV$^2$) are shown in each case.
References are: 
(a,b)\protect\cite{cool,akimov}; (c,d)\protect\cite{schamberger}. 
Smaller $s'$-values are not shown
in order to avoid distortions due to resolution.
In order to minimize non-\pom -exchange background, only points with
$\xi < 0.03$ are plotted. 
}
\label{fig:snorm1}
\end{center}
\end{figure}

Figures~\ref{fig:snorm1}(a-d) show this quantity 
from three fixed-target experiments at Fermilab\cite{schamberger,akimov,cool},
plotted vs. $s'$ ($s' = \xi s$) in 4 different \T -bins;
only data with $s' > 3$~GeV$^2$ 
and $\xi < 0.03$ are plotted, in order to avoid 
resolution smearing on the low side, and background contributions on the high
side.
In each case, there are data at several different $s$-values.

We may draw the following conclusions from Figs.~\ref{fig:snorm1}.
There appears to be pronounced resonance structure, 
whose \T -dependence is steeper than for the general \pom -exchange signal.
There are striking dependences on $s$, mainly at the lowest $|t|$-values,
thus limiting the validity of
Eq.~\ref{eq:tripleR} for $s'$ values less than about 12~GeV$^2$
and for $s$ below about 400~GeV$^2$.

\section{Longitudinal structure in diffraction}
\label{longi}
\indent

Longitudinal energy-flow along the beam direction (``beam-jets'')
is a well-established property of high energy
hadronic interactions.
One of the dominant features of diffractive interactions that has come
to light in recent years is that the outgoing particle distributions in the 
diffractive center-of-mass ($X$ system) are also $p_t$-limited and 
far from the isotropic shape expected from the old
``fireball'' model of diffraction. 
Since ``beam jets'' tend to contain spectator-like partons from the beam
particles, one may speculate that
the observation and study of such jets in diffractive systems might
eventually lead to further understanding of the partonic structure of the \pom .

Early evidence for $p_t$-limited distributions in diffraction is found
in Refs.~\cite{ptlim1,ptlim2,ptlim3,ptlim4}.
Experiment R608 at the CERN Intersecting Storage Rings
studied several exclusive diffractive final
states from React.~\ref{eq:dif} and 
found that the final-state $X$ systems possess sharp longitudinal 
structures\cite{blois1,pomqrk,p4pi} in their centers-of-mass.
The UA4 Collaboration\cite{ua4lim},
using measured track pseudorapidity distributions at $\sqrt{s} = 546$~GeV, 
also presented evidence against the ``fireball'' model of diffractive 
systems.

One of the most striking examples of longitudinal structure 
is in the following reaction from the R608 experiment\cite{blois1,pomqrk}:
\begin{equation}
p \, p \, \ra \, ( \lamb \alamb p ) \, p, 
\label{eq:pomqrk}
\end{equation}
\noindent
where (\lamb \alamb p) is the $X$ system and
all three particles are seen in the final state.
In the $X$ center-of-mass, a forward (backward) \lamb\ is 
always correlated with a backward (forward) proton.
The particles both have average momenta, normalized to \rsp /2,
of $|0.6|$.
For both cases, the \alamb\ is at rest (its average momentum is 0.0).

\begin{figure}
\begin{center}
\mbox{\epsfig{file=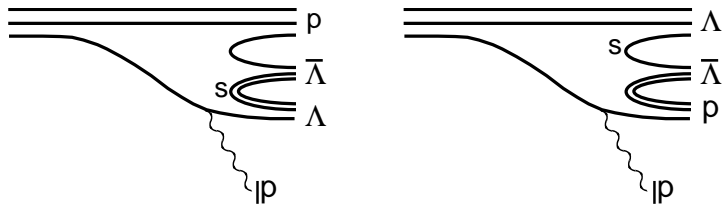,width=10cm}}
\end{center}
\caption[]{
The diagrams show the dominant \pom -proton interaction processes for
React.~\protect\ref{eq:pomqrk} which correspond to the two peaks
seen in Refs.~\protect\cite{blois1,pomqrk}. 
}
\label{fig:pomqrk}
\end{figure}

Figure~\ref{fig:pomqrk} shows the diagrams which correspond to the 
reported effect.
In each case the exchanged \pom\ appears to interact
with a single quark in the incident proton and to ``kick'' it backwards,
thereby stretching the color string between quark and diquark. 
The observed baryon pair production in this final state requires that
both a diquark pair and a quark pair be produced. Depending on their
relative spatial ordering, the lambda (proton) is forward or backward
(backward or forward). 
In either case the \alamb\ remains essentially at rest in the
$X$ system. The R608 results 
were interpreted as evidence for \pom -single quark interactions\cite{pomqrk}.

In Refs.~\protect\cite{blois1,pomqrk}, data on the final state 
$p \ra \lamb \phi K ^+$ are also presented. In that case, 
two $s \overline{s}$
quark pairs come from the stretched string, the \lamb\ usually goes
forward and the $\phi$ remains at rest in the $X$ system.

Experiment R608 also studied the reaction: 
$p \ra \lamb K^+$ \cite{lund,henkes}. Not only are
the \lamb\ and K$^+$ sharply peaked forward and backward, respectively,
but the \lamb\ is observed to have a very large polarization, over 60\%.
This remarkably large polarization is due in large part to the
absence of dilution effects from
$\Sigma ^{*}$ and $\Sigma ^{\circ}$ decay background.

Because the \rsp\ values in the reactions studied in R608 were
rather small (e.g. 3 GeV), compared with the enormously larger values
in the present UA8 experiment, we may expect that the details of the 
\pom -proton interaction will be quite different.

\subsection{Energy-flow}
\label{flow}
\indent

With the use of the calorimeter information associated with each event in
our data sample of React.~\ref{eq:dif}, we look for
evidence of longitudinal structure in the inclusive flow of energy in the
center-of-mass of the diffractive system.
Since, for the events studied here, the calorimeter was not used in the
trigger, its offline use 
allows for an unbiased study of the $X$-system in the events.

We define an energy-flow quantity, $dE_{CM}/d(cos \theta_{CM})$, where 
$E_{CM}$ and $\theta_{CM}$ are the energy
and polar angle of a particular UA2 calorimeter cell,
in the diffractive center-of-mass.
This quantity has the advantage that isotropy in the energy-flow 
would appear as a flat distribution in $cos \theta_{CM}$.

\begin{figure}
\begin{center}
\mbox{\epsfig{file=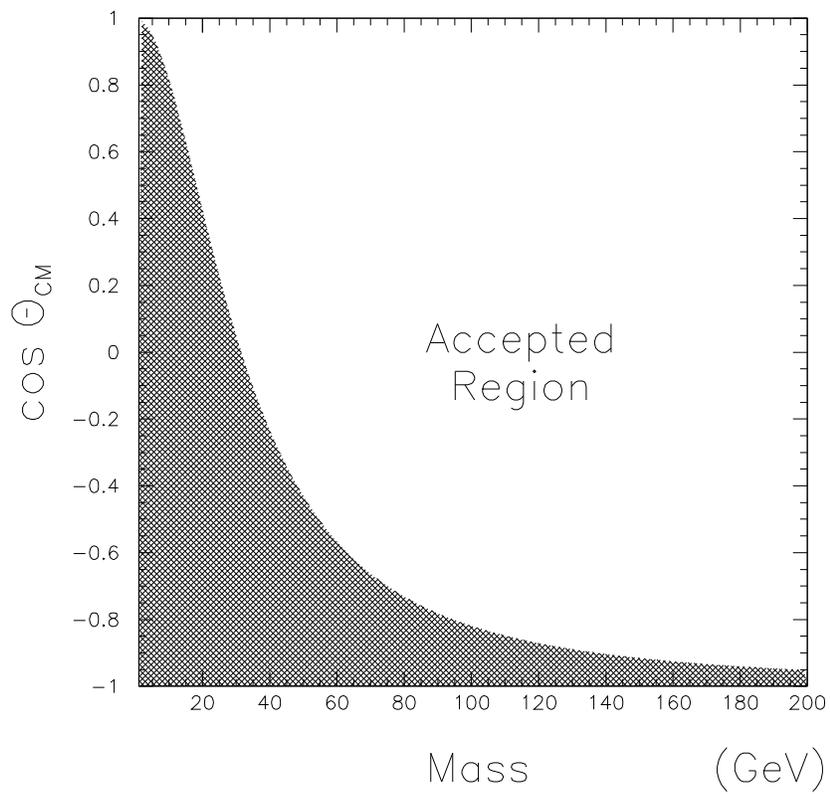,width=12.7cm}}
\end{center}
\caption[]{
$cos(\theta_{CM})$ acceptance range in the UA2 calorimeter
as a function of the mass of the diffractive system. 
The unshaded region shows the range which is accepted. The
asymmetry in acceptance is due to the motion of the diffractive 
center-of-mass in the laboratory.
$cos(\theta_{CM}) > 0$ is the \pom\ hemisphere.
}
\label{fig:flowac}
\end{figure}

In order to obtain $dE_{CM}/d(cos \theta_{CM}$) we note that,
at fixed $s'$, each UA2~$\theta$--cell boundary transforms to a
unique $\rm \theta_{CM}$ value.
Thus the total energy in each laboratory $\theta$-bin in the UA2 calorimeter
can be transformed to a corresponding
$\theta_{CM}$-bin in the $X$ center-of-mass.
Figure~\ref{fig:flowac} shows the range of $\rm \theta_{CM}$ ``seen'' by the UA2 
calorimeter system as a function of diffractive mass.
The motion of the $X$ center-of-mass is responsible
for the forward-backward asymmetry in the acceptance in this system
and for the fact that, for diffractive masses larger than about 40~GeV,
we have almost perfect acceptance in the \pom\ hemisphere.

\begin{figure}
\begin{center}
\mbox{\epsfig{file=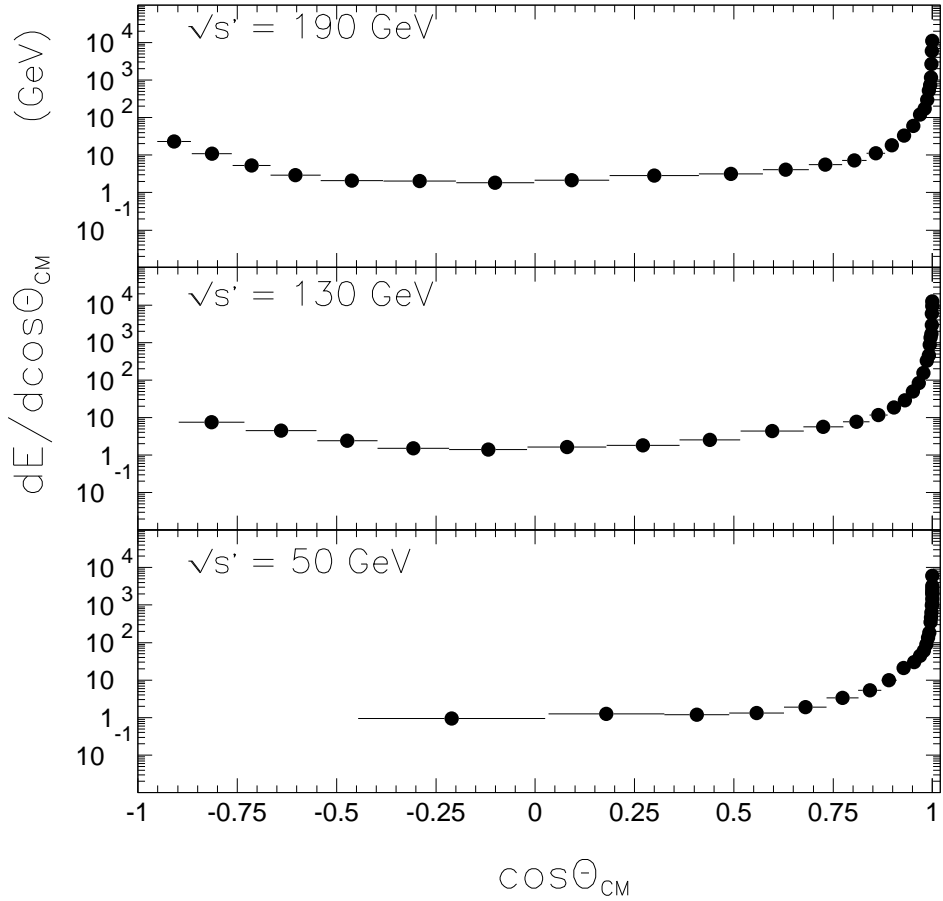,width=12.7cm}}
\end{center}
\caption[]{
Average energy flow per event in the diffractive center-of-mass
for three mass bins. 
Each point corresponds to the set of UA2 calorimeter cells with a common 
$\rm\theta_{LAB}$.
In constructing the plot, the energy, and $\rm \cos (\theta_{LAB})$
from each set of cells, are transformed into 
the diffractive center-of-mass.
On each plot, $ \cos (\theta_{CM}) > 1$ corresponds to 
the \pom\ hemisphere.
The striking peaks at $ \cos (\theta_{CM}) = \pm 1$ are 
characteristic of a 
\peetee -limited longitudinal event structure.

}
\label{fig:dedcos}
\end{figure}

Table~\ref{tab:vdedcos} contains the average values (per event) of
$ dE_{CM}/d(cos \theta_{CM})$ vs. $\theta_{CM}$
for \rsp\ values of 50, 130 and 190 GeV, respectively.
Figures~\ref{fig:dedcos}(a-c) show these values plotted vs. $\theta_{CM}$.
It is clear that the energy-flow in the center-of-mass of diffractive systems
is highly anisotropic and that there is a sharp peak along the \pom\ 
axis, which may be interpreted as
a \pom\ ``beam jet'' in the \pom -proton interaction.
The essential shape for $\rsp < 200$ GeV is observed to be independent of $s'$,
(and also of whether or not jets are present in the final 
state --- not shown here).
From the results in Fig.~\ref{fig:dedcos}, coupled with the R608 results,
we may infer that a similar sharp peak also exists in the proton direction.

\section{Conclusions}
\label{conclude}
\indent

We have presented the results of a measurement of
the differential cross section
for the single-diffractive reactions:
$p_i + \ap \, \ra \, p_f + X$ \, and \, 
$p + \ap _i  \, \ra \, X + \ap_f$
at \rsps , in the momentum transfer range, $0.8 < -t < 2.0$~GeV$^2$
and final state Feynman-$\xp > 0.90$. 

Double differential cross sections are compared
with previous measurements at the ISR and \SPS -Colliders. 
Despite the order of magnitude increase in center of mass energy 
and the large differences in total cross sections, 
we find only a small difference in diffractive 
cross sections from lower energy measurements.

Several new features of the \pom\ Flux Factor are determined from
simultaneous fits to these UA8 data and similar data from the CHLM 
collaboration at the CERN-ISR.
Prominent among these is that the \pom\ Regge trajectory requires a term 
quadratic in \T , with coefficient, $\alpha '' = 0.079 \pm 0.012$~GeV$^{-4}$,
which may indicate the onset of the hard \pom , or a contribution from
multiple \pom -exchange.  

We also show that
existing data require a \pom -proton cross section that decreases at small
mass and increases at large mass, similar to other reactive
total cross sections.
We have obtained the parameter, 
$K \sigma _0 = 0.72 \pm 0.10$~mb/GeV$^2$, where $K$ is the unknown
normalization constant of the \pom\ Flux Factor in the proton and $\sigma _0$
is the constant in the \pom -proton total cross section at high energies,
$\sigma _0 (s')^{\epsilon}$. The data are inconsistent with
the $s$-dependent $K$ proposed by Goulianos.

Finally, we used the UA2 calorimetry to investigate the energy-flow in the 
diffractive final state.
A striking longitudinal structure is observed in the final state which 
supplies further evidence that the \pom\ undergoes a hard interaction with 
one or more partons of the target particle.

\section{Acknowledgements}

We thank John Collins and Peter Landshoff for helpful discussions.
We are grateful to the CERN administration for their support,
and also to the UA2 collaboration, without whose continuing
help and cooperation, the calorimeter measurements would not have been possible.
A.K. and N.O. wish to thank Cukurova University, TUBITAK and 
ICSC-World Lab for support. For the latter, they are particularly grateful
to Professors A. Zichichi and T. Ypsilantis. 

\pagebreak

\pagebreak

\begin{table}
\centering
\vspace{2ex}
\begin{tabular}{|l|c|} \hline
Selection Cuts&Remaining Fraction\\ \hline
Triggers                        &1.0 \\
One Track with $\xp > 0.9$  &0.59 \\                          
No Hit in Adjacent Spectrometer &0.58 \\                          
Pileup                          &0.52 \\                          
Halo                            &0.45 \\                          
Energy Topology (see text)      &0.42 \\ \hline                   
\end{tabular}
\caption[]{Effect of offline cleanup cuts on the initial triggered data sample
of 150K events. The fraction of the initial data sample which remains after
each cut is shown.
The final combined data sample for both Roman-pot positions 
(12$\sigma$ and 14$\sigma$, respectively, from the beam orbit) 
is 62,627 events. 
The YN and NN events (see text) 
are rejected in the final ``event topology'' cut. 
See the text for explanations of the various cuts.
}
\label{tab:select}
\end{table}   

\begin{table}
\centering
\vspace{1ex}
\begin{tabular}{|l|r|} \hline
Trigger Component or Offline Cut &Efficiency \\ \hline
Trigger Processor: chamber hit logic    & 0.902 \\
Trigger Processor: momentum calculation & 0.974 \\
Scintillation Veto: halo                & 0.995 \\ 
Scintillation Veto: adjacent arm        & 0.998 \\
Off-line Cut: pileup (average)           & 0.908 \\
Off-line Cut: halo                       & 0.916 \\ \hline
Net Efficiency (not including geometric acceptance) & 0.726 \\ \hline
\end{tabular}
\vspace{1ex}
\caption[]{The various Trigger and Cut Efficiencies used in cross section 
calculations.
The geometric acceptance is not included in this table.
}
\label{tab:efficac}
\end{table}

\begin{table}
\centering
\vspace{ 2ex}
\begin{tabular}{|l|c c c c c c c|} \hline
     & & &-$t$ & &(GeV$^2$) & & \\
     &0.95          &1.05          &1.15          &1.30
                    &1.50          &1.70          &1.90 \\
$\xi$ & & & & & & & \\ \hline
0.09 & 0.77$\pm$0.08 & 0.65$\pm$0.08 & 0.38$\pm$0.06 & 0.26$\pm$0.02
                     & 0.14$\pm$0.02 & 0.11$\pm$0.02 & 0.05$\pm$0.02 \\
0.08 & 0.79$\pm$0.07 & 0.53$\pm$0.06 & 0.43$\pm$0.06 & 0.24$\pm$0.02
                     & 0.16$\pm$0.02 & 0.13$\pm$0.02 & 0.06$\pm$0.02 \\
0.07 & 0.66$\pm$0.07 & 0.44$\pm$0.06 & 0.43$\pm$0.06 & 0.30$\pm$0.02
                     & 0.17$\pm$0.02 & 0.09$\pm$0.02 & 0.04$\pm$0.02 \\
0.06 & 0.65$\pm$0.07 & 0.58$\pm$0.07 & 0.49$\pm$0.06 & 0.23$\pm$0.02
                     & 0.17$\pm$0.02 & 0.11$\pm$0.02 & 0.10$\pm$0.02 \\
0.05 & 0.70$\pm$0.07 & 0.50$\pm$0.06 & 0.42$\pm$0.06 & 0.27$\pm$0.02
                     & 0.16$\pm$0.02 & 0.11$\pm$0.02 & 0.08$\pm$0.02 \\
0.04 & 0.81$\pm$0.07 & 0.60$\pm$0.07 & 0.48$\pm$0.06 & 0.29$\pm$0.02
                     & 0.17$\pm$0.02 & 0.14$\pm$0.02 & 0.09$\pm$0.02 \\
0.03 & 0.99$\pm$0.08 & 0.59$\pm$0.06 & 0.45$\pm$0.06 & 0.37$\pm$0.03 
                     & 0.20$\pm$0.02 & 0.16$\pm$0.02 & 0.08$\pm$0.02 \\
\hline
\end{tabular}    
\caption[]{Single Diffractive Differential Cross Sections: \dsig\ (mb/GeV$^2$).
The cross sections are single arm and correspond to 
$p \ap \ra p + X$ \, \, OR \, \, $p \ap \ra \ap + X$.
The errors shown are statistical only. Not included is a 10--15 \% systematic
uncertainty in the absolute cross section. See text for a discussion of
the uncertainty in the \T -values.
}
\label{tab:sigmatm}
\end{table}

\begin{table}
\centering
\vspace{ 2ex}
\begin{tabular}{|c|c|}  \hline
$-t$ (GeV$^2$) &$d \sigma / dt$  (mb/GeV$^2$) \\ 
\hline
0.85 &0.2060 $\pm$ 0.0036 \\
0.95 &0.1520 $\pm$ 0.0031 \\
1.05 &0.1040 $\pm$ 0.0016 \\
1.15 &0.0855 $\pm$ 0.0015 \\
1.25 &0.0608 $\pm$ 0.0013 \\
1.35 &0.0437 $\pm$ 0.0011 \\
1.45 &0.0339 $\pm$ 0.0010 \\
1.55 &0.0261 $\pm$ 0.0009 \\
1.65 &0.0218 $\pm$ 0.0009 \\
1.75 &0.0168 $\pm$ 0.0008 \\
1.85 &0.0135 $\pm$ 0.0007 \\
1.95 &0.0092 $\pm$ 0.0006 \\ 
\hline
\end{tabular}
\caption[]{Single Diffractive Differential Cross Sections: 
$d \sigma/dt$~~(mb/GeV$^2$) for $\xp > 0.95$. 
The cross sections are single arm and correspond to 
$p \ap \ra p + X$ \, \, OR \, \, $p \ap \ra \ap + X$ (single arm).
The points are plotted in Fig.~\ref{fig:dsigdt}.
The integral for the $|t|$ region, 1--2 GeV$^2$, is 41.5~$\mu$b.
}
\label{tab:dsigdt}
\end{table}


\begin{table}
\centering
\vspace{ 2ex}
\begin{tabular}{|c|c|c|c|c|c|c|} 
\hline 
$-t$  &$\chi^2$  &$\chi^2$/D.F.  &$\alpha (t)$
     &$\chi^2$  &$\chi^2$/D.F.  &$\alpha (t)$   \\ 
bin  & & &1-comp. & & &2-component \\
\hline
1.0--1.1 &19.0 &2.7  &$1.01 \pm 0.01$  &17.0 &  2.4 &$0.94 \pm 0.01$ \\
1.1--1.2 &15.5 &2.2  &$0.98 \pm 0.01$  &14.3 &  2.0 &$0.91 \pm 0.01$ \\
1.2--1.4 &2.4  &0.3  &$0.98 \pm 0.01$  &6.5  &  0.9 &$0.92 \pm 0.01$ \\
1.4--1.6 &3.1  &0.4  &$0.93 \pm 0.01$  &6.9  &  1.0 &$0.89 \pm 0.01$ \\
\hline
\end{tabular}
\caption[]{
Fit results for the \pom\ Regge trajectory in four bins of \T , from
1.0--1.6 GeV$^2$. In each case results are given for two types of fits,
the first assuming 1-component in \sigpomtot\ and the second assuming
2-components in \sigpomtot . In each case, the $\chi ^2$ and the
$\chi ^2$ per degree of freedom are given.
}
\label{tab:mvalues}
\end{table}

\begin{table}
\centering
\vspace{ 2ex}
\begin{tabular}{|r c||c|c|c|c|} \hline
$\chi ^2$  &$\chi ^2$/DF &$K \sigma _0$ (mb GeV$^{-2}$)  
          &$\alpha ''$ (GeV$^{-4}$) &$b$ (GeV$^{-2}$)  &$R$ \\ 
\hline 
413 &8.8 &0.68 $\pm$ 0.02 &--                &--             &--           \\
391 &8.5 &0.86 $\pm$ 0.05 &-0.029 $\pm$0.006 &--             &--           \\
331 &7.2 &1.28 $\pm$ 0.08 &--                &0.62 $\pm$0.06 &--           \\
199 &4.4 &2.32 $\pm$ 0.15 &0.156 $\pm$0.011  &2.4 $\pm$0.2   &--           \\
108 &2.3 &0.51 $\pm$ 0.02 &--                &--             &5.2 $\pm$0.4 \\
98  &2.2 &0.40 $\pm$ 0.03 &--                &-0.20 $\pm$0.06&6.3 $\pm$0.6 \\
85  &1.9 &0.41 $\pm$ 0.03 &0.020 $\pm$0.004  &--             &6.4 $\pm$0.5 \\
65  &1.5 &0.67 $\pm$ 0.08 &0.078 $\pm$0.013  &0.88 $\pm$0.19 &5.0 $\pm$0.6 \\
\hline
\end{tabular}    
\caption[]{
Fit results of Eqs.~\ref{eq:tripleR} and \ref{eq:twocomp} to
experimental values of \dsig\ (mb/GeV$^2$)
in the $\xi$-range, 0.03--0.04.
There are 48 data points in all fits.
}
\label{tab:chisq}
\end{table}

\begin{table}
\centering
\vspace{ 2ex}
\begin{tabular}{|l|c|c|c|c|} 
\hline 
                      &Fit ``A"     &Fit ``B"     &Fit ``C"     &Fit ``D"   \\
\hline
$\chi ^2$             &65           &117          &357          &393          \\
Data                  &48           &84           &257          &292          \\
$\chi ^2$/DF          &1.5          &1.5          &1.4          &1.4   	      \\
$K \sigma _0$ (mb GeV$^{-2}$)  
                       &$0.67\pm0.08$   &$0.53\pm0.14$  
                                              &$0.81\pm0.13$   &$0.72\pm0.10$\\
$\alpha ''$ (GeV$^{-4}$)
                      &$0.078\pm0.013$ &$0.055\pm0.012$
                                             &$0.074\pm0.014$ &$0.079\pm0.012$\\
$b$  (GeV$^{-2}$)      &$0.88\pm0.19$   &$0.66\pm0.18$ 
                                              &$0.95\pm0.24$   &$1.08\pm0.20$ \\
$R$                     &$5.0\pm0.6$     &$6.8\pm0.7$ 
                                               &$2.8\pm0.5$     &$4.0\pm0.6$  \\
\hline 
A(ua8)                &--           &$23\pm8$     &--            &$25\pm7$    \\
A(551)                &--           &--           &$296\pm40$   &$280\pm30$   \\
A(930)                &--           &--           &$232\pm26$   &$226\pm21$   \\
c(ua8)~~~(GeV$^{-2})$ &--          &$2.2\pm0.2$   &--           &$2.1\pm0.2$  \\
c(isr)~~~~(GeV$^{-2})$ &--           &--          &$3.5\pm0.2$  &$3.5\pm0.1$  \\
\hline
\end{tabular}    
\caption[]{Fit results of Eqs.~\ref{eq:tripleR} and \ref{eq:twocomp} to
experimental values of \dsig\ (mb/GeV$^2$).
Fit ``A" is to both UA8 and ISR data in the $\xi$-range, 0.03--0.04.
Fits ``B", ``C" and ``D" add the indicated UA8 and/or ISR 
data in the $\xi$-range, 0.04--0.09, to the Fit ``A" sample,
and include background in the fit of the form, A$\xi^1 e^{ct}$,
as discussed in the text. 
}
\label{tab:chisqbknd}
\end{table}


\begin{table}
\centering
\begin{tabular}{|l c|l c|l c|} 
\hline 
 & & & & & \\
 \multicolumn{2}{|c|}{\rsp = 50  \, GeV} 
& \multicolumn{2}{c|}{\rsp = 130 \, GeV}
& \multicolumn{2}{c|}{\rsp = 190 \, GeV} \\
 & & & & & \\
$cos(\theta_{CM})$ & ${{dE_{CM}}\over{dcos(\theta_{CM})}}$ &
$cos(\theta_{CM})$ & ${{dE_{CM}}\over{dcos(\theta_{CM})}}$ &
$cos(\theta_{CM})$ & ${{dE_{CM}}\over{dcos(\theta_{CM})}}$ \\ 
 & & & & & \\
\hline
-0.211   &1.04$\pm$0.02  &-0.815 &7.88$\pm$0.16   &-0.909 &23.05$\pm$0.50\\
-0.179   &1.36$\pm$0.02  &-0.640 &4.57$\pm$0.09   &-0.814 &11.36$\pm$0.25\\
-0.407   &1.30$\pm$0.02  &-0.473 &2.44$\pm$0.06   &-0.714 & 5.34$\pm$0.14\\
0.558   & 1.44$\pm$0.03   &-0.307&1.58$\pm$0.04   &-0.603 & 2.86$\pm$0.09\\
0.680   & 2.06$\pm$0.04   &-0.118&1.48$\pm$0.04   &-0.461 & 2.12$\pm$0.06\\
0.774   & 3.69$\pm$0.07   &0.080 &1.78$\pm$0.05   &-0.291 & 2.09$\pm$0.06\\
0.842   & 5.98$\pm$0.12   &0.271 &1.91$\pm$0.05   &-0.101 & 1.93$\pm$0.06\\
0.890   & 10.8$\pm$0.3    &0.440 &2.63$\pm$0.08   &0.092  & 2.22$\pm$0.07\\
0.927   & 22.5$\pm$0.4    &0.597 &4.48$\pm$0.09   &0.300  & 2.90$\pm$0.07\\
0.954   &   33$\pm$1     &0.725  &5.83$\pm$0.14   &0.493  & 3.17$\pm$0.09\\
0.969   &   49$\pm$1     &0.808  &7.96$\pm$0.22   &0.631  & 3.92$\pm$0.11\\
0.979   &   70$\pm$2     &0.864  &12.3$\pm$0.4    &0.730  & 5.57$\pm$0.18\\
0.985   &  102$\pm$3     &0.903  &18.9$\pm$0.6    &0.803  & 7.41$\pm$0.25\\
0.989   &  152$\pm$5     &0.931  &29.8$\pm$0.9    &0.857  & 11.4$\pm$0.4\\
0.993   &  209$\pm$7     &0.951  &50.1$\pm$1.5    &0.898  & 18.7$\pm$0.7\\
0.995   &  375$\pm$11    &0.966  &87.4$\pm$2.5    &0.929  & 34.2$\pm$1.0\\
0.997   &  495$\pm$16    &0.977  &161 $\pm$4      &0.952  & 62.2$\pm$1.7\\
0.998   &  724$\pm$24    &0.985  &330 $\pm$8      &0.969  &  124$\pm$3\\
0.9986  &  930$\pm$53    &0.991  &487 $\pm$17     &0.980  &  182$\pm$7\\
0.9991  & 1260$\pm$60    &0.994  &885 $\pm$25     &0.986  &  318$\pm$10\\
0.9994  & 1450$\pm$90    &0.996  &1400$\pm$40     &0.991  &  530$\pm$17\\
0.9996  & 1540$\pm$120   &0.997  &1720$\pm$60     &0.994  &  745$\pm$24\\
0.9997  & 4650$\pm$250   &0.9980 &2730$\pm$100    &0.996  & 1230$\pm$40\\
0.9998  & 2160$\pm$220   &0.9987 &5530$\pm$180    &0.997  & 2550$\pm$70\\
0.9999  & 2190$\pm$260   &0.9992 &9230$\pm$240    &0.998  & 5820$\pm$130\\
0.99995 & 3190$\pm$420  &0.9996 &12900$\pm$410    &0.999  &11230$\pm$250\\ 
\hline
\end{tabular}        
\caption[]{The average Energy-Flow per event (GeV) in the 
center-of-mass of the $X$ System
in React.~\protect\ref{eq:dif}: 
$dE_{CM}/dcos(\theta_{CM})$ for three different values of \protect\rsp .
In each case, the 26 entries correspond to the 26 $\theta$-cells of the
UA2 calorimeter (8 in each end-cap plus 10 in the central calorimeter).
The $\theta_{CM}$ values are quoted at the center of each cell.
}
\label{tab:vdedcos}
\end{table}

\clearpage
 
\clearpage

\clearpage

\clearpage

\clearpage

\clearpage

\clearpage

\clearpage

\clearpage

\clearpage

\clearpage

\clearpage

\clearpage

\clearpage

\clearpage

\clearpage

\clearpage

\clearpage

\clearpage

\clearpage

\clearpage

\clearpage

\clearpage

\clearpage

\clearpage

\clearpage

\clearpage

\clearpage

\clearpage

\clearpage

\clearpage

\clearpage

\clearpage

\clearpage

\clearpage

\clearpage

\end{document}